\documentclass{sig-alternate}
\usepackage{mathptmx} 

\usepackage{fancyhdr}
\usepackage[normalem]{ulem}
\usepackage[hyphens]{url}
\usepackage[sort,nocompress]{cite}
\usepackage[final]{microtype}
\usepackage[keeplastbox]{flushend}
\usepackage[bookmarks=true,breaklinks=true,letterpaper=true,colorlinks,citecolor=blue,linkcolor=blue,urlcolor=blue]{hyperref}

\newcommand{\red}[1]{\textcolor{red}{#1}}

\newcommand{\projectname}{RCP\xspace}
\newcommand{\ise}{ISE\xspace}

\usepackage{multirow}
\usepackage{comment}
\usepackage{xspace}

\usepackage{subfig}
\usepackage{wrapfig}

\usepackage{mathdesign}
\usepackage{amsmath}
\usepackage{pifont}

\newcommand{\todo}[1]{\textcolor{black}{#1}}

\newcommand{\rev}[1]{\textcolor{black}{#1}}

\author{
  You Wu\\
  \texttt{University of Southern California}\\
  \texttt{Email:youwu@usc.edu}\\
  \texttt{alchem.usc.edu}
  \and
  Xuehai Qian\\
  \texttt{University of Southern California}\\
  \texttt{Email:xuehai.qian@usc.edu} \\
  \texttt{alchem.usc.edu}
}

\title{\projectname: A Low-overhead Reversible Coherence Protocol} 

\begin{document}
\maketitle
\pagestyle{plain}


\begin{abstract}

This paper proposes \projectname, a new reversible 
coherence protocol that ensures invisible 
speculative load execution (ISLE) with low overhead. 
\projectname can be combined with 
processor mechanisms that eliminate the 
effects of speculative instructions on other 
instructions to achieve low overhead
{\em invisible speculative execution (ISE)}.
ISE provides protection that is at least as strong as 
speculative privacy tracking (SPT) and stronger than 
speculative taint tracking (STT). 
\projectname is designed by systematically
extending the existing 
coherence protocol to incorporate speculative 
loads and states. 
The protocol is implemented in Gem5 and verified
with Murphi. 
The results show that \projectname based ISE
incurs lower overhead than STT/SDO/SPT while
providing similar/stronger protection.

\end{abstract}

\section{Introduction}
\label{intro}

As Spectre \cite{kocher2019spectre}, Meltdown\cite{lipp2018meltdown} and other attacks demonstrated, modern processor architectures based on speculation are facing major security issues. These attacks exploit speculative execution to modify or leave traces in the cache hierarchy, and extract secretes using side-channels.
With several years of intensive research, 
various kinds of vulnerabilities are identified.
It is critical that the defense mechanisms are able to 
provide comprehensive protection.

\begin{figure}[t]
    \centering
    \includegraphics[width=\linewidth]{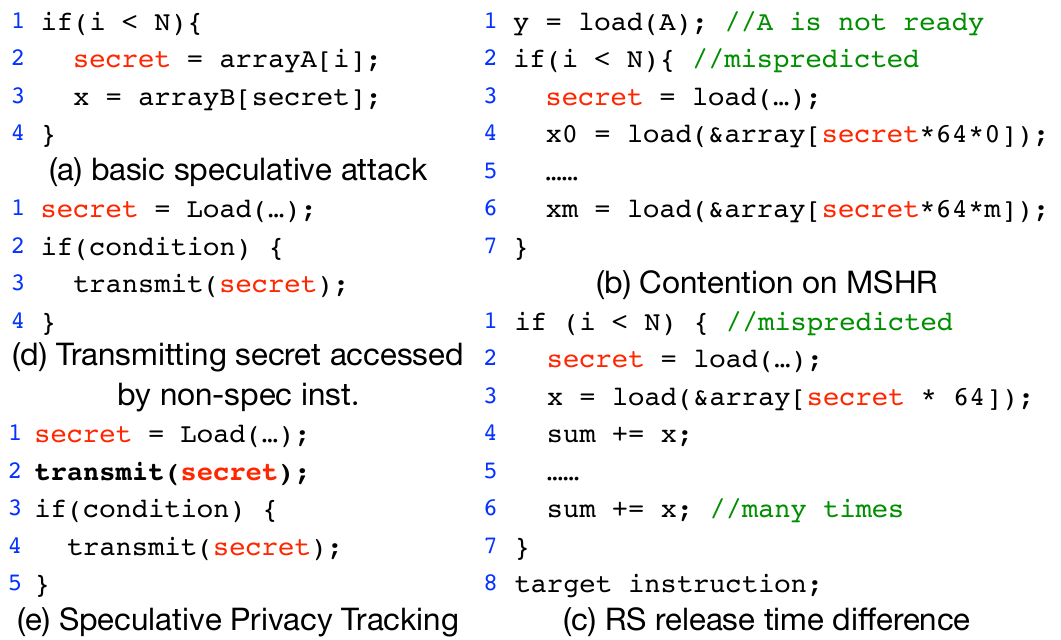} 
    \vspace{-4mm}
    \caption{Speculative Execution Attacks}
    \vspace{-8mm}
    \label{attack_summary}
\end{figure}

Figure~\ref{attack_summary} shows four typical attacks. 
For the {\em basic attack} (a), an attacker can exploit branch
misprediction to leak secret (or arbitrary memory) via the
data cache. It first primes the branch to predict 
that the condition is true by making the code repeatedly
run with valid value of \texttt{i}.
Then the attacker provides an out-of-bound value for \texttt{i}
and the processor (mis)predicts that the condition
is still true and speculatively loads out-of-bound data
depending on the secret. 
Once the processor resolves the misprediction, 
it rolls back execution but the data accessed persists in 
cache. 
The secret can be extracted by common techniques, e.g., 
Flush+Reload, that exploit the cache side-channel vulnerabilities.

Figure~\ref{attack_summary} (b) and (c) show
two advanced {\em speculative interference attacks}~\cite{behnia2021speculative}. 
In (b), the non-speculative load(A) is not 
ready to execute and the processor executes 
the speculative instructions after the branch. 
If the secret is 1, the sequence of loads will 
occupy all MSHR entries. 
When load(A) is ready,
it will be delayed until the branch is resolved 
because there is no available MSHR entry. 
If the secret is 0, all speculative loads
only occupy one MSHR entry since they access the 
same address, and load(A) can be issued without 
delay. 
In (c), if secret is 1, the load (line 4)
will miss, and 
the execution of the following dependent 
speculative instructions
will occupy all reservation station (RS) entries.
When the load returns, the following instructions
will be dispatched, and the target instruction 
is executed after the branch is resolved. 
If secret is 0, the load hits in cache, the 
following instructions can be dispatched and 
executed faster. Thus the target instruction can be
executed before the branch is resolved. 
In both cases, the secret is encoded in the execution timing
of the non-speculative instruction. 
Figure~\ref{attack_summary} (d) abstracts
a general attack that accesses the secret with a 
non-speculative instruction, and transmits 
it in speculative execution. 

Speculative execution is an important 
performance enhancing technique used in all commercial
processors. 
The architects understood the needs and mechanisms
for rolling back the execution and recovering
the {\em architectural (processor) states} when 
the speculative execution is squashed. 
However, {\em no systematic mechanisms are in place to 
reverse the 
state changes due to speculative execution in 
cache hierarchy}. 
We believe that it is one of the crucial causes (but not all)
of the speculative attacks. 
The goal of the paper is to close this gap. 
We first briefly review the existing defense
schemes.

\begin{figure*}[t]
    \centering
    \includegraphics[width=\linewidth]{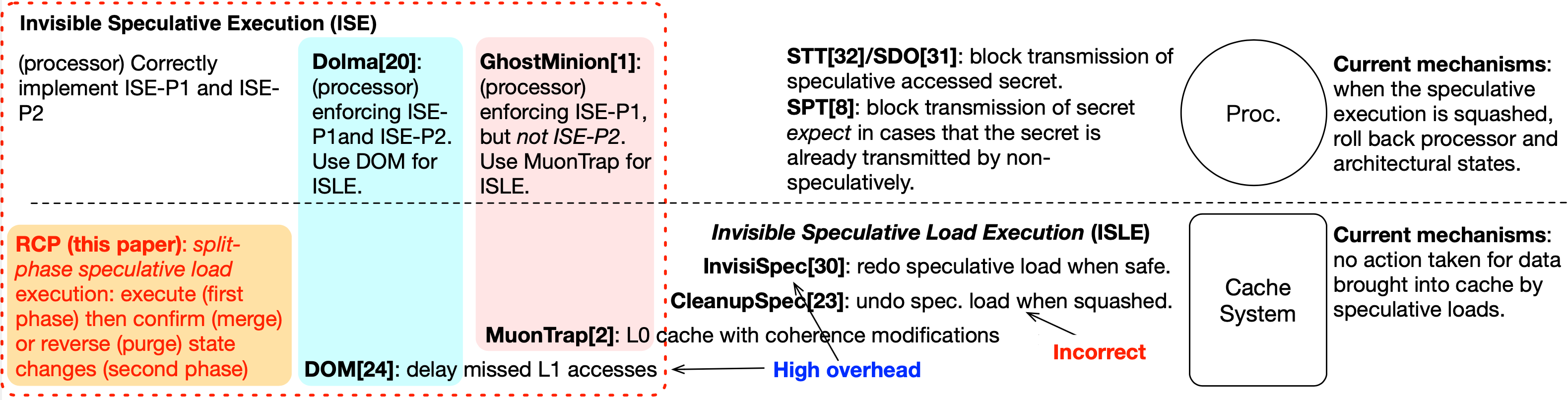}
    \caption{\rev{The Key Contribution: {\em Reversible Coherence Protocol} and Its Relation to Other Schemes} }
    \label{solution_compare}
\end{figure*}

{\bf ISLE: incomplete solution focusing on cache. }
Early mitigation mechanisms~\cite{ainsworth2020muontrap,khasawneh2019safespec,saileshwar2019cleanupspec,yan2018invisispec,sakalis2019efficient} focused on 
{\em invisible speculative load execution (ISLE)}, i.e., ensuring that the executed but then squashed speculative loads 
do not leave trace in cache hierarchy. 
Although ISLE can defend specific
attacks such as Spectre similar to the basic attack in Figure~\ref{attack_summary} (a),
it cannot provide comprehensive protection
since it does not consider other speculative 
instructions.
Thus, it cannot defend speculative interference attacks
in Figure~\ref{attack_summary} (b) and (c). 
It can defend the attack in 
Figure~\ref{attack_summary} (d) if the transmission
is through cache.

\rev{While multiple ISLE schemes have been proposed, 
the prior works did not explicitly 
present the correctness criteria of the ISLE schemes, making it 
possible to overlook subtle design errors that cause the speculative load
execution not completely invisible. 
This paper takes the {\em first} step to define 
the properties in Section~\ref{sec_formal}
based on the principle that {\em unsafe} speculative loads should not 
affect the state and location of a cache block, and as a result, the timing
of non-speculative loads/stores accessing the block.}
\rev{Based on these properties, }
we discovered that the low-overhead
scheme CleanupSpec~\cite{saileshwar2019cleanupspec} 
does not correctly implement ISLE, leaving security vulnerabilities. 
\rev{InvisiSpec~\cite{yan2018invisispec} and DOM~\cite{sakalis2019efficient} are correct but come
with higher overhead. 
MuonTrap~\cite{ainsworth2020muontrap} is an effective scheme, it is correct and incurs low overhead.
As discussed later, we consider our proposed scheme as an alternative 
with an interesting difference.}

{\bf STT/SDO/SPT: (mostly) comprehensive processor-stall based
solutions with higher overheads. }
Yu {\em et al.}~\cite{yu2019speculative}
took a systematic approach and 
recognized three key steps for a 
successful speculative execution based
attack: speculatively {\em accessing} the secret,
{\em sending} it through microarchitectural covert channels, and {\em receiving} the secret.
Based on this insight, 
speculative taint tracking (STT) 
(NDA~\cite{weisse2019nda} uses similar strategy)
is developed with the principle of 
{\em blocking the transmission of secret}:
allowing the access of secret but
stalling execution of 
secret-dependent instructions.
A key feature of this approach is that {\em the 
cache design and coherence protocol are not modified}.

STT provides stronger security property than ISLE and 
can defend attacks in
Figure~\ref{attack_summary} (b) and (c)
from ~\cite{behnia2021speculative} (similar to 
SpectreRewind~\cite{fustos2020spectrerewind}).
For Figure~\ref{attack_summary} (b),
STT defends such attack by stalling
all speculative loads because they all depend on the secret.
For Figure~\ref{attack_summary} (c),
STT defends the attack by stalling 
all secret dependent instructions.
The reported results indicate that 
the overhead of STT is high, i.e., 
$\sim$14.5\%, which is reduced by the 
optimized design SDO~\cite{yu2020speculative} to $\sim$11\%.

The design principle of STT/SDO makes them
unable to defend the attack in Figure~\ref{attack_summary} (d).
They only block speculative 
transmission of speculatively accessed secrets, 
but {\em do not block the transmission of 
non-speculatively accessed secrets}.
A straightforward modification
is to treat all accessed data as secrets, 
taint them, and block the dependent speculative 
execution. 
But it will likely block most 
speculative execution, making the 
performance unattractive. 
Speculative Privacy Tracking (SPT)~\cite{choudhary2021speculative} reuses
the taint tracking mechanisms in STT and achieves
lower overheads by not blocking a speculative secret transmitter 
if the secret is already transmitted non-speculatively.
With SPT, the speculative transmission instruction
in Figure~\ref{attack_summary} (e) is not blocked. 
Nevertheless, SPT still blocks more speculative execution than STT,
and introduces even higher overheads (45\% vs. 14.5\%).

\rev{\bf Our approach: {\em processor-stall+ISLE with reversible 
coherence protocol}. }
Based on the \rev{discussion so far}, neither ISLE 
nor processor-stall based solutions
is optimal: ISLE is not comprehensive while STT/SDO/SPT incur
higher overheads. 
In this paper, we take a {\em hybrid} approach:
{\em minimally stalling the processor to eliminate the 
interference of speculative execution on other instructions; and
designing a correct and low-overhead reversible
cache coherence protocol for ISLE}.
This approach is inspired by~\cite{behnia2021speculative}, which
demonstrated the insufficient protection of ISLE,
and proposed additional processor mechanisms for eliminating the effects of speculative instructions on other instructions.

\begin{figure*}[t]
    \centering
    \includegraphics[width=\linewidth]{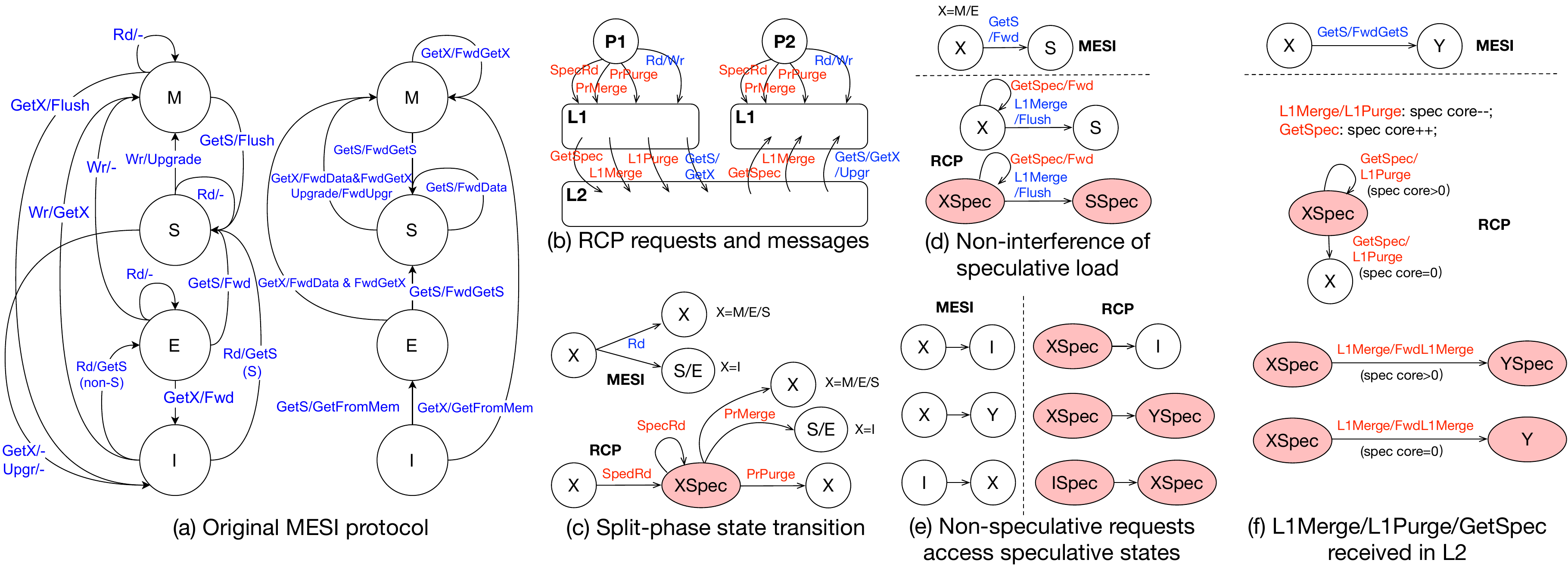}
    \vspace{-6mm}
    \caption{Key Insights of Reversible Coherence Protocol}
    \label{insight}
    \vspace{-6mm}
\end{figure*}

We call this approach as
{\em Invisible Speculative Execution (\ise)}.
\rev{According to~\cite{behnia2021speculative}}, 
\ise provides two strong {\em security properties}:

$\bullet$ {\bf \ise-SP1}: the set of instructions committed and their timing
should not be influenced by speculative execution;

$\bullet$ {\bf \ise-SP2}: the time for squashing a set of instructions should not depend on speculatively accessed data.

The two properties can be realized by enforcing three
requirements with {\em implementation mechanisms}:

\rev{$\bullet$ {\bf ISLE}: it ensures the security properties 
{\bf \ise-SP1} and {\bf \ise-SP2}
for {\em speculative loads}}. We achieve this goal by proposing the systematically
designed {\bf Reversible Coherence Protocol (\projectname)} with low overhead;

$\bullet$ {\bf \ise-P1}: ``{\em no instruction ever influences the execution time
of an older instruction}''~\cite{behnia2021speculative}; and 

$\bullet$ {\bf \ise-P2}: ``{\em any resources allocated to an instruction at the interface
of the frontend and the execution engine are not deallocated until
the instruction becomes non-speculative}''~\cite{behnia2021speculative}.

\rev{{\bf \ise-P1} and {\bf \ise-P2} ensure 
{\bf \ise-SP1} and {\bf \ise-SP2} for {\em speculative non-load instructions}.} 
In a nutshell, the relationship between \ise and ISLE
is: 

\begin{center}
\framebox{\todo{\bf \ise=\ise-P1+\ise-P2+ISLE}}
\end{center}

We do not claim 
the processor policies {\bf \ise-P1} and {\bf \ise-P2},
or the general \ise approach
as our contribution since they were outlined in~\cite{behnia2021speculative}.
The {\bf key contribution} of the paper is the novel coherence
protocol to implement ISLE. 

\todo{\ise can defend the speculative interference
attacks.
For Figure~\ref{attack_summary} (b), 
{\bf \ise-P1} ensures that an occupied MSHR entry for a speculative load
is released and allocated for Load(A) when
it is ready to execute. This can be implemented
by preemption mechanism. 
For Figure~\ref{attack_summary} (c), 
{\bf \ise-P2} ensures that, even the speculative
instructions (the many additions) can finish 
execution faster if (secret=0), their 
RS entries cannot be deallocated until the
branch is resolved. 
\ise---oblivious to 
the source of the dependent secret---provides {\em stronger protection than STT/SDO} 
and prevents attacks in Figure~\ref{attack_summary} (d).}We 

{\bf Relation to alternative \ise schemes:} Figure~\ref{solution_compare} show the relations between \ise and all existing 
solutions. \rev{Unlike the early ISLE schemes, \ise provides
comprehensive protection---even stronger than 
STT/SDO and similar to SPT. There are two alternative \ise schemes}. 
GhostMinion~\cite{ainsworth2021ghostminion} rephrases {\bf \ise-SP1} as
``Strictness Ordering'' and {\bf \ise-P1} as ``Temporal Ordering'', which 
is the basis of its implementation. 
However, it does not correctly 
implement \ise for two reasons:
(1) it does not enforce {\bf \ise-P2} \rev{therefore cannot
defend the attack in Figure~\ref{attack_summary} (c)}; and 
(2) it uses MuonTrap~\cite{ainsworth2020muontrap} as the mechanism for ISLE, 
which is not fully correct based on the security properties in Section~\ref{sec_formal}. 
Dolma~\cite{loughlin2021dolma} adopts processor mechanisms
similar to {\bf ISE-P1} and {\bf ISE-P2}
to ensure transient non-observability by 
STT-like speculative information flow tracking.
Dolma uses Delay On Miss (DOM) to support ISLE, which allows speculative execution if the instruction hits in L1 and delays that on miss. 
Our results show that DOM incurs higher overheads 
than \projectname, {\em which can be also applied to 
Dolma}.
\rev{Thanks to \projectname, our \ise scheme incurs low overhead 
than STT/SDO/SPT while providing similar or stronger protection}.

{\bf Key insights of \projectname:} 
To design \projectname, we take a principled 
approach, and show that it can be done 
by extending the existing coherence protocol in a 
systematic manner. 
\rev{The key property of \projectname is:
{\em when an unsafe speculative load is squashed, it should be as if it is not 
executed; when it is later committed, it should behave like a normal load
in the ordinary coherence protocol}.
The second part of the property is interesting: it is {\em not} 
a correctness property but merely a design choice. 
InvisiSpec~\cite{yan2018invisispec} and DOM~\cite{sakalis2019efficient} 
satisfy this with high overhead, while MuonTrap~\cite{ainsworth2020muontrap} does not satisfy this,
i.e., a safe speculative load can behave different than a normal load,
but still correct. }
We demonstrate the design of \projectname
by extending MESI protocol for two level caches (private L1 and 
shared L2) as shown in Figure~\ref{insight} (a).
\rev{The complexity of 
\projectname is relatively high, but our implementation in 
a simulator is fully functional and can run comprehensive 
benchmarks to completion. We are confident that our 
implementation can be used as an initial reference design 
when the protocol is adopted in a real implementation. 
We will also make our code publicly available}.
Specifically, \projectname is developed based on 
the following key insights, which 
can be also applied to designs based on other protocols
and cache organizations. 

$\bullet$ {\bf Introducing new requests and messages
for speculative load}.
\projectname introduces three new operations in the 
interface between processor and 
cache coherence, and allow them to
participate in coherence operations:
(1) speculative load $SpedRd$; 
(2) $PrMerge$, which is performed when a speculative load becomes safe; and 
(3) $PrPurge$, which is performed when a speculative load is squashed. They lead to additional messages:
(a) $GetSpec$, the forwarded speculative load request
to the owner; (b) $L1Merge$, the message from L1 to L2 and 
other L1s when a speculative load is merged; and
(c) $L1Purge$, the message from L1 to L2 (but not to 
other L1s) when a speculative load is purged.
The new (in red) and existing (in blue)
requests and messages in a two-processor
setting are illustrated in 
Figure~\ref{insight} (b). 
In \projectname, the transitions for requests
other than the speculative load are exactly the same as
Figure~\ref{insight} (a).

$\bullet$ {\bf Split-phase state transition}.
With merge and purge, the state transitions for a 
speculative load are split into multiple phases. 
\rev{As shown in Figure~\ref{insight} (c),
when a $SpecRd$ is issued, the cache line state $X$ transitions
to the corresponding {\em speculative state} $XSpec$.
Later, if the $SpecRd$ becomes safe, i.e., $PrMerge$ is received, 
$XSpec$ transitions to the corresponding non-speculative state;
otherwise, i.e., $PrPurge$ is received, $XSpec$ transitions back to $X$.
The key problem is to specify the transitions among speculative states.}

$\bullet$ {\bf Non-interference of speculative load}.
In ISLE, a $SpecRd$ should not cause state changes to 
remote caches since it will incur timing changes. 
\rev{In \projectname, when a $SpecRd$ misses in L1, $GetSpec$ is forwarded to the
current owner, which just provides the data response to $GetSpec$
without changing state. 
The state change is finalized later when the $SpecRd$ becomes 
safe (merged).} The insight is shown in Figure~\ref{insight} (d).
In comparison, both CleanupSpec and MuonTrap stall
the $GetSpec$.

$\bullet$ {\bf Mechanical extension of state transition graph}.
\rev{In \projectname, two scenarios can lead to multiple state transitions among 
speculative or non-speculative states:
(1) when non-speculative requests access cache
lines in speculative states; and 
(2) when $L1Merge$ are $L1Purge$ are
received in shared L2.}
The state transitions can be designed 
by mechanically extending the original MESI protocol. 
For (1), the insight is that the non-speculative requests will 
trigger state transitions in the same manner as in
MESI but among corresponding speculative states. 
For (2), whether a state should
transition from speculative to non-speculative
depends on whether there is still
speculatively accessed copies in any L1.
\rev{We maintain a {\em spec core} counter
for each cache line indicating the number
of speculative copies in L1.
This information can be obtained efficiently with counting bloom filter (see
Section~\ref{specBuf}).
We design the state transitions of \projectname in L2
based on the current status of spec core. 
The insights are shown in Figure~\ref{insight} (e) and (f), and 
the detailed transitions are discussed in Section~\ref{spec_trans}.}

{\bf Advantages of \projectname.}
The coherence actions triggered by 
the merge and purge operations are {\em not}
in the critical path of the execution and 
can be performed in the cache hierarchy 
{\em concurrently} with processor execution.
\rev{It explains why the additional traffic due to merge and purge
does not lead to high performance overhead}. 
in our results, 
even with certain traffic overheads than InvisiSpec,
the performance overheads of \projectname is 
lower.
Importantly, the requests and messages related to 
ordinary loads/stores and speculative load
are handled by \projectname in a {\em unified} manner. 
It means that \projectname is not likely 
to leak side-channel
information on when speculative loads become safe or when
squashes happen. 
The recent work~\cite{mengming_li_2021_5771649} indeed showed the possibility of 
leaking the information on whether 
squash has happened due to 
CleanupSpec's potentially secret dependent squash operations.

{\bf Evaluation highlights.} We implemented \projectname in 
Gem5~\cite{binkert2011gem5}. 
The correctness is verified using Murphi model checking tool~\cite{276232}.
We compare the overheads in two scenarios:
(1) using \projectname for ISLE, 
comparing to InvisiSpec, CleanupSpec and DOM;
(2) using \projectname as the ISLE mechanism for \ise ({\em \ise-\projectname}),
comparing to using InvisiSpec for ISLE ({\em \ise-IS}). We also compare the two \ise designs with STT/SDO/SPT.
Both {\em \ise-\projectname} and {\em \ise-IS}
offer stronger security property than STT/SDO, and 
comparable property to SPT.  
For (1), our results show that \projectname incurs slowdown of 
(7.7\%,7.4\%) on (SPEC2017,PARSEC) under TSO,
lower than (12.6\%,18.3\%) for InvisiSpec,
(9.3\%,24.2\%) for CleanupSpec. DOM incurs considerably higher overheads
compared to others. 
For (2), we show that {\em \ise-\projectname} incurs slowdown of 
(8.4\%,11\%) on (SPEC2017,PARSEC) under TSO,
lower than (12.7\%,18.7\%) for {\em \ise-IS}.
The results under RC show similar trends.
STT and SDO incurs
(22.5\%,32.2\%), (10.9\%, 15.1\%) overheads on (SPEC2017,PARSEC),
respectively; 
while SPT incurs much higher slowdown $\sim$ 48.5\% on SPEC2017.
We observe that the overheads due to 
processor mechanisms to prohibit interference
from speculative instructions is very low.

\section{Background}
\subsection{Out-of-Order Execution}
Modern processors perform
speculative out-of-order
execution to exploit instruction
level parallelism. 
Recent studies~\cite{koruyeh2018spectre,lipp2018meltdown} revealed that the secret-dependent
speculative execution
can cause irreversible cache
state changes or data movements (e.g.,
left the speculatively accessed cache
block in L1 cache) that can 
lead to a covert channels.
It is because the speculative load is secret-dependent. 
\subsection{Threat Model}
\label{threat}

We focus on defending transient attacks
enabled by speculative {\em instructions}, rather than speculative loads.
We assume the
SameThread and Cross-Core models and do not 
consider simultaneous multithreading (SMT), which can be prevented by recent techniques such as adding defense on context switches~\cite{ainsworth2020muontrap} or making the cache way-partitioned to avoid SMT-side channels\cite{saileshwar2019cleanupspec}. 
Similar to all existing works~\cite{saileshwar2019cleanupspec,yan2018invisispec,yu2020speculative,yu2019speculative}, we do not consider 
speculative stores. 
In a typical design, a store buffer sits between processor 
and L1 cache, and a store starts performing---sending invalidations---after
it is retired from reorder buffer (ROB) and inserted to the store buffer.
Thus, SpectrePrime and MeltdownPrime~\cite{trippel2018meltdownprime} which leverage
speculative stores are out-of-scope.
Our evaluation implements a processor model which does not 
send invalidations until instruction retirement. 
The recent papers~\cite{ren2021see,kim2021uc,deng2021leaky} proposed a new Spectre variant which exploits the micro-op cache at the frontends to create the new timing side-channel. This vulnerability can be mitigated by delaying the update of micro-op cache until the micro-op eventually commits.  
This variants of spectre attacks are also out-of-scope.
We assume that attackers can 
measure the latency of loads
and stores, but cannot measure the 
latency change due to increased coherence traffic. 
It is implicitly assumed by existing ISLE schemes
since they
all incur additional traffic. 
\rev{Specifically, additional traffic is generated for correctly speculated loads
in InvisiSpec, due to the redo operation; and MuonTrap, due to additional 
operation to ensure the consistency of coherence states,
e.g., asynchronous upgrades in Figure~\ref{fig:exist_problems}.
It is generated for incorrectly speculated loads in CleanupSpec for the 
undo operation. 
In comparison, \projectname generates traffic for both correctly
speculated and squashed speculative loads. 
We acknowledge that the additional traffic may be explored as a side-channel but
no such attack has been demonstrated. 
Also, it is a potential vulnerability for all schemes. In fact, \projectname
may leak less information compared to others since the traffic is incurred
for all speculatively executed loads. }

Our threat model 
includes the attacks that change the timing of committed instructions before
or after mis-speculation, such as SpectreRewind~\cite{fustos2020spectrerewind} and 
Speculative Interference~\cite{behnia2021speculative}. 
During mis-speculation duration, depending on the secret, 
different sequences of speculative instructions are executed---putting
different pressure on non-speculative instructions before or after
mis-speculation duration. 

\subsection{Existing Defense Schemes}
\label{sec:exist}

{\bf ISLE schemes.} InvisiSpec~\cite{yan2018invisispec} 
issues a speculative load without affecting cache states and 
a second load when it becomes safe,
which will change cache state and leave trace in 
cache hierarchy. 
Sakalis {\em et al.}
\cite{sakalis2019efficient,tran2020clearing} delays speculative load on L1 miss (DOM) and avoids processor stall by value prediction. 
A speculative load does not change the coherence states when missing in the L1 cache. 
DOM is simple solution, but our results show that
it incurs considerably higher overheads than InvisiSpec. 
CleanupSpec~\cite{saileshwar2019cleanupspec} is an undo approach with low overheads.
On mis-speculation, besides squashing execution effects within the processor, 
the cache system performs cleanup operations to invalidate or roll back to the state before the mis-speculation.
MuonTrap~\cite{ainsworth2020muontrap} introduces an L0 filter cache and restrict coherence operations
to limit speculative changes to a small region. 
Later, we show that \rev{CleanupSpec does not correctly 
implement ISLE}. 


{\bf Processor-stall based schemes. }
STT/SDO~\cite{yu2019speculative,yu2020speculative} tracks 
speculative accessed data and 
blocks secret-dependent execution. 
SPT~\cite{choudhary2021speculative} blocks speculative transmission
of non-speculatively accessed secret only
if it is not transmitted non-speculatively.

{\bf Pattern-based Schemes: }
Conditional speculation~\cite{sakalis2019efficient} defines 
security dependency and stalls speculative 
execution based on the patterns.
SpecCFI~\cite{koruyeh2020speccfi} performs static analysis
on control flow graph to prevent the 
malicious indirect branch.




\section{ISLE Security Properties}
\label{sec_formal}

\begin{figure*}[t]
    \centering
    \includegraphics[width=\linewidth]{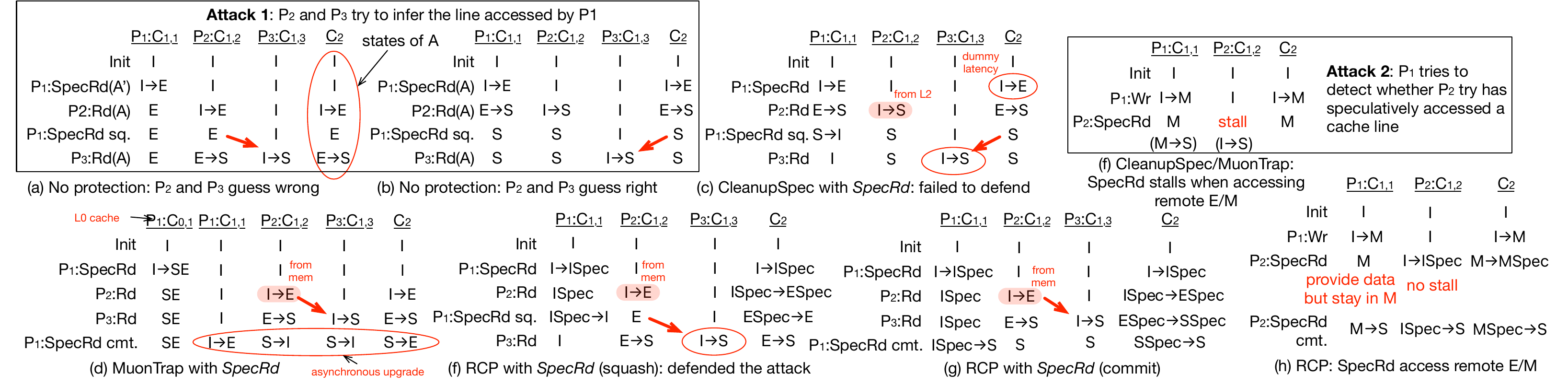}
    \vspace{-6mm}
    \caption{\rev{CleanupSpec~\cite{saileshwar2019cleanupspec} and MuonTrap~\cite{ainsworth2020muontrap}, compared with \projectname}}
    \vspace{-6mm}
    \label{fig:exist_problems}
\end{figure*}


\subsection{Definitions}
\label{sec_def}

A speculative load ($SpecRd$) has three key points during its execution:
1) {\em Issue point (I)}: when it
is ready to be issued speculatively;
2) {\em Non-speculative point (NS)}: when it
is no longer speculative, either becoming
safe or squashed; and 
3) {\em Globally perform point (G)}: 
when all effects of $SpecRd$
are finalized---either becoming a part
of system state (if it becomes safe at NS),
or completely cleared (if it is 
squashed at NS)---and the write which produces the 
read value has been globally performed. 
We assume atomic writes, so that G
for write is well-defined. 
Non-atomic writes issues are discussed in Section~\ref{sec_analysis}.
$[I,G]$ is called pending period. 
If a $SpecRd$ becomes
safe and committed {(denoted as a property C for 
each $SpecRd$)}, we have
$SpecRd[C]=true$, otherwise $SpecRd[C]=false$.

We consider two 
{\em ``effects''} of a $SpecRd$:
1) {\em locations} of the cache line A
after G, defined as a function $L[A]$; 
and 2) {\em states} of the cache line
after G, defined as a function $S[A]$. 
Each $L_i[A]$ is a bit vector 
$e_{i,1}, ... , e_{i,m}$, where
$m$ is the number of private caches in level $i$; or a bit $e_i$ when the level $i$ is shared. 
Each $e_{i,j}, j \in \{1,...,m\}$ or $e_i$
indicates whether the cache line
resides in the corresponding cache.
$L[A]$ is a concatenated
bit vector indicating present information of a cache line
among cache components in all levels.
For $S[A]$, $e_{i,j}$ or $e_i$ are replaced with 
$s_{i,j}$ or $s_i$, which indicates
the cache line state.
If the pending period of $a \in \{Rd,Wr\}$ is 
not overlapped with any other accesses'
pending periods, 
$L(\mathcal{L},a)[A]$ and $S(\mathcal{L},a)[A]$
define
the locations and states of the cache line 
containing address $A$ after $G_{a}$,
starting from the initial 
locations $\mathcal{L}$ and $\mathcal{S}$ before 
the execution of $a$.

To capture the effects of a number of concurrent
accesses $(a_1,...,a_n)$, $a_i \in \{Rd,Wr\}$,
$L(\mathcal{L},(a_1,...,a_n)_{[I_{min},G_{max}]})[A]$ and $S(\mathcal{S},(a_1,...,a_n)_{[I_{min},G_{max}]})[A]$
specify the locations and states of the cache line
after all accesses are globally performed
serialized with {\em certain} total order.
$I_{min}$ and $G_{max}$ are the minimum
and maximum I and G among $a_1,...,a_n$.
Each access may cause the changes
of the locations and states. 
Assuming the total order is 
$a_1 \rightarrow a_2 \rightarrow ... \rightarrow a_n$, the 
sequences of the location and state
change are 
$\mathcal{L}_{a_1} \rightarrow \mathcal{L}_{a_2} \rightarrow ... \rightarrow \mathcal{L}_{a_n}$ and
$\mathcal{S}_{a_1} \rightarrow \mathcal{S}_{a_2} \rightarrow ... \rightarrow \mathcal{S}_{a_n}$.
The transition functions
when $a_i$ is serialized right after $a_{i-1}$
are: $L(\mathcal{L}_{a_{i-1}},{a_i})[A]=\mathcal{L}_{a_{i}}$ and
$S(\mathcal{S}_{a_{i-1}},{a_i})[A]=\mathcal{S}_{a_{i}}$.


\subsection{Security Property}
\label{sec_prop}


\rev{{\bf Property 1: (Non-overlapping) No effects from mis-speculated load.}}
Consider a $SpecRd$, $[I_{SpecRd},G_{SpecRd}]$ is not overlapped with any other access.
If $SpecRd[C]=false$, the 
locations and states of the cache line
should be equivalent to the execution 
before $SpecRd$:
$L(\mathcal{L},SpecRd)[A]=L_{G_{SpecRd}}[A]=L_{I_{SpecRd}}[A]$.
The states can be handled similarly by replacing $\mathcal{L}$ and $L$
with $\mathcal{S}$ and $S$.

\rev{{\bf Remark}: For the correct speculated load, the 
same effects as a non-speculative load is {\em not} a required security property.
A coherence protocol can exhibit different behaviors for non-speculative load
and a speculative load that later becomes safe. 
MuonTrap~\cite{ainsworth2020muontrap} is one such example.}

{\bf Property 2: (Overlapping) Serialization
of the correct speculated loads.}
Consider two accesses with 
overlapping pending period, 
the first is a speculative load $SpecRd$,
the second is an access $acc \in \{Rd,SpecRd,Wr\}$.
If $SpecRd[C]=true$, then 
the two accesses should be correctly
serialized. From Property 1, the locations
and states of the cache line should be equivalent to
the execution that replaces the 
speculative loads with the corresponding non-speculative load:
$L((SpecRd,acc))[A]=L(L(\mathcal{L},SpecRd),acc)$, 
if $SpecRd$ $\rightarrow$ $acc$; 
$L((SpecRd,acc))[A]=L(L(\mathcal{L},acc),SpecRd)$, 
if $acc$ $\rightarrow$ $SpecRd$.
Equations for states are similar. 

{\bf Property 3: (Overlapping) No effects 
to overlapping accesses from mis-speculated loads. }
If $SpecRd[C]=false$, 
the squashed $SpecRd$ should not have 
any effects on other overlapped accesses:
$L(\mathcal{L},(SpecRd,acc))[A]=L(\mathcal{L},acc)[A]$. 
Equation for states is similar. 

{\bf Corollary: A non-speculative request 
$acc \in \{Rd,Wr\}$ is not aware of any 
overlapping speculative $SpecRd$ to the
same address before NS point of $SpecRd$}. It can be directly
obtained from Property 3.
If the execution is affected by $SpecRd$ before its NS point
and $SpecRd$ is later squashed, 
$acc$ may not reach $L(\mathcal{L},acc)[A]$
and $S(\mathcal{S},acc)[A]$.

\subsection{Existing ISLE Solutions and \projectname}
\label{analyze_exist}


{\bf InvisiSpec~\cite{yan2018invisispec} and DOM~\cite{sakalis2019efficient}: \rev{correct with high overhead}}.
They satisfy all security properties by
not allowing the speculative load
to participate the coherence protocol
before NS point.

{\bf CleanupSpec~\cite{saileshwar2019cleanupspec}: \rev{incorrect}}.
Consider a sequence of memory accesses
from three processors shown in Figure~\ref{fig:exist_problems} (a) and (b), $P_1$ is the victim, $P_2$ and $P_3$ are
attackers. 
$P_1$
is induced to speculatively access a cache line $A$ (but later squashed) whose address will reveal secret.
$P_2$ tries to guess and access a cache line that may be $A$
before $P_1$'s $SpecRd$ is squashed, then
and $P_3$ can infer the line by 
measuring the response latency difference caused
by {\em irreversible cache state changes}.
Specifically, if the guess is wrong, $P_3$ gets
the line from $C_{1,2}$, which 
gets the forwarded request from $C_2$, since $P_2$ is the first to access the line (Figure~\ref{fig:exist_problems} (a));
if the guess is correct (both $P_1$ and $P_2$ have 
accessed the line), 
$P_3$ will directly get the line from $C_2$ (indicated as the red arrow) with shorter latency since $P_1$ is the first
to access the line and $P_2$ has changed it to shared (Figure~\ref{fig:exist_problems} (b)). 
Due to different latency, $P_3$ can infer whether the 
guess is correct, and if so, the line accessed by $P_1$.
Our security properties prevent such attack
by guaranteeing, even if the line is speculatively 
accessed by $P_1$, if the load is squashed, 
$P_3$ should get the line with the {\em same} latency as if $P_1$
did not execute the $SpecRd$.
A protocol satisfying these 
properties will behave in the same way as Figure~\ref{fig:exist_problems} (a).

CleanupSpec's behavior for the 
``correct guess'' case is
illustrated in Figure~\ref{fig:exist_problems} (c).
CleanupSpec allows the state changes to $E$ in $C_{1,1}$,
when $P_2$ accesses it before $P_1$'s $SpecRd$ is squashed, $C_{1,1}$ is the one who forwards the line.
It is different from the scenario if $P_1$'s $SpecRd$ does
not exist, in which $P_2$ should experience a cache 
miss in both L1 and L2 and get the data from memory.
Thus, the $SpecRd$ effects the timing of $P_2$'s request
(violation of Property 3).
In an attempt to ensure the property, 
CleanupSpec adds a ``dummy latency'' by forcing an artificial 
cache miss
in $C_{2}$, which effectively
prevents $P_2$ to sense the latency difference. 
It ensures Property 3 for $SpecRd$ w.r.t. $P_2$'s 
request. 
Later, when $P_1$'s $SpecRd$ is squashed, 
$C_{1,1}$ locally invalidates the line. 
Unfortunately, this operation does not reverse the state change in $C_2$.
As a result, when later $P_3$ accesses the cache line,
it will get the response from $C_2$ with a shorter latency since it is in shared state (both $P_1$ and $P_2$ have accessed it). But the attacker knows that the latency should be longer---forwarded by $C_{1,2}$---if $P_1$ had not accessed the line. 
This example explains that CleanupSpec does not protect
such an attack and {\bf Property 3 is violated} 
by $SpecRd$ in $P_1$ w.r.t. $P_3$'s request:
the mis-speculation in $P_1$ effects 
the state of the line in $C_2$ (making it shared), 
which is later inferred by $P_3$ with 
response latency difference.

{\bf MuonTrap~\cite{ainsworth2020muontrap}: \rev{correct with low overhead}}.
It uses an L0 cache 
to keep the speculatively accessed data
in restrictive cases 
when coherence state changes are not 
exposed. 
The basic design degrades an MESI protocol to MSI.
To recover the benefit from $E$ state, 
MuonTrap introduces $SE$ state in L0. 
A line is brought to L0 as $SE$ by $SpecRd$
when a $Rd$ would have brought it to L1 
as $E$. It ``behaves like S to the coherence protocol'', but when the $SpecRd$ is 
committed, an ``asynchronous upgrade'' is performed to invalidate the other copies so 
that the line can be install in L1 as $E$.
Figure~\ref{fig:exist_problems} (d) shows 
how the previous example works in MuonTrap.
$P_2$ and $P_3$ have the 
same behavior as no $SpecRd$ from $P_1$, thanks to the L0. 
\rev{In contrast, when $P_1$'s $SpecRd$ becomes safe, it 
causes different effects than a normal $Rd$:
the asynchronous upgrade operation
makes every 
line brought into L0 cache by (safe) $SpecRd$ behave like writes.
Even though the behavior is different, MuonTrap is still a correct ISLE
implementation with low overhead. 
It is a smart idea to not require the same behavior as a non-speculative load 
so that the design of the protocol is more flexible. }

\rev{{\bf{\projectname}: correct with low overhead}.
\projectname satisfies the three properties and an {\em additional 
``overcapture'' requirement}: for the correct speculated load, 
\projectname exhibits the {\em same} behavior as a non-speculative load. 
In a sense, \projectname is correct but more conservative than MuonTrap. 
We made the design choice because enforcing the property is more natural 
in our framework. 
We believe \projectname is still an important contribution because
(1) the design is completely different from all prior schemes;
(2) it further makes the correct and incorrect speculation indistinguishable.
While there is no attack leveraging the difference between correct and incorrect
speculation, but a scheme that close such difference may avoid potential future
vulnerabilities more easily. }

\rev{It is worth noting that a design difference does affect performance overhead.}
Both CleanupSpec and MuonTrap stall a 
forwarded request of $SpecRd$ 
when it is about to change
a remote L1 cache copy from $M/E$ to $S$, shown
in Figure~\ref{fig:exist_problems} (e).
\projectname avoids such stall with the non-interference property. 
The behaviors of \projectname for Attack 1\&2 are shown in 
Figure~\ref{fig:exist_problems} (f)(g)(h), which will
be discussed in 
Section~\ref{sec_analysis}.

\section{Hardware Structure}

\subsection{Processor Supports}
\label{model}

The processor tracks Visibility Point (VP) dynamically, determined by the attack model. 
In the Spectre-model, an instruction reaches
VP if all older control-flow instructions have resolved. 
In the Futuristic-model, an instruction reaches
VP if it cannot be squashed for any reason. 
All the instructions before (after) VP are considered to be unsafe (safe).
With VP maintained during execution, 
the process can determine whether each instruction
becomes safe in each cycle. 
When an instruction is fetched, it is marked
as ``unsafe''. When the load is issued, 
if it becomes safe, then a $Rd$ is generated;
otherwise, a $SpecRd$ is issued.
The update of VP in a cycle 
will trigger the merge or purge of sequence of 
instructions.
To purge a sequence of speculative loads,
only the $PrPurge$ for the oldest one is sent and all younger 
ones are squashed together.
For merge, only the $PrMerge$ for the the youngest is sent and all older ones
will be merged. 
The recent Pinned Load work~\cite{zhao2022pinned} proposed
optimization to speed-up the advance of the VP to either
reduce the stall time in processor-stall based solutions,
or allow safe loads to be issued earlier. 
The idea is applicable to \projectname and
allows $PrMerge$ to be issued earlier---further reducing the overheads. 

To enforce {\bf \ise-P1}, each instruction 
is assigned a priority tag based on the program order when inserted into ROB. 
An older instruction has higher priority than a younger instruction. 
When an instruction is about to be stalled due to
shared resource contention (e.g., MSHR or EU), 
the incoming instruction's priority tag
is compared to the tags of instructions that have occupied the resource.
If some instructions have lower priority, the one with the lowest priority
(thus the youngest instruction) is preempted, and will be re-scheduled
in the next cycle. 
To enforce {\bf \ise-P2}, when an instruction become safe or squashed, the ROB
informs instruction queue to deallocate the shared resource
allocated to the instruction. Thus, for a squashed instruction, the 
shared resource is only deallocated when at the squash point, but not earlier. 

\subsection{Speculative Buffer Structure}
\label{specBuf}

\projectname uses speculative buffer (specBuf) to keep the 
effects of speculation. 
Similar to InvisiSpec~\cite{yan2018invisispec}: there is a 
one-to-one mapping relation between a processor's load queue (LQ)
\setlength{\intextsep}{2pt}%
\setlength{\columnsep}{8pt}%
\begin{wrapfigure}[7]{r}{0.5\linewidth}
    \centering
    \vspace{-1mm}
    \includegraphics[width=\linewidth]{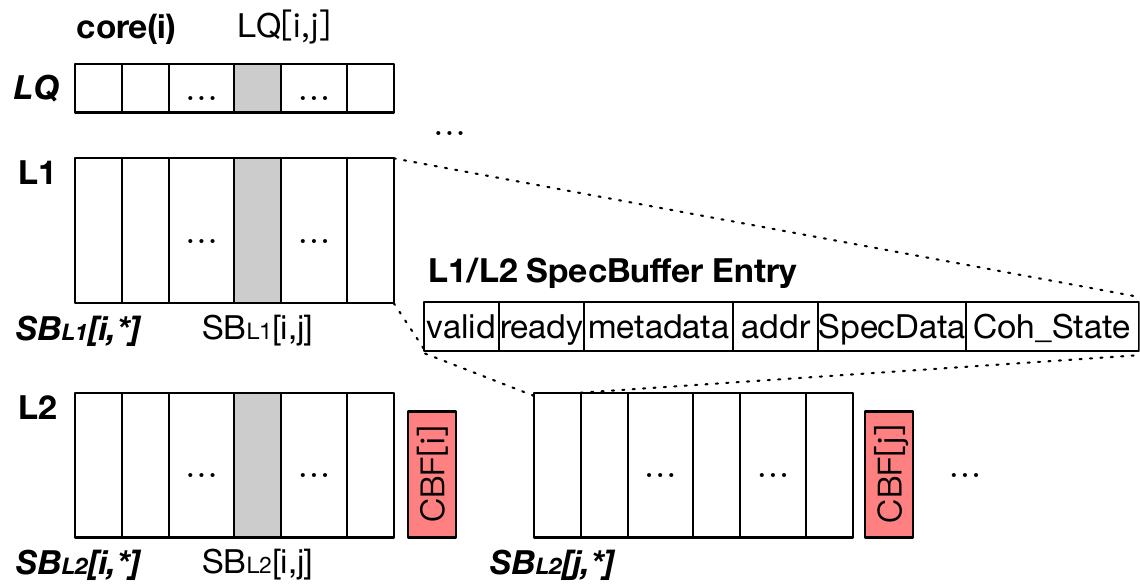}
    \vspace{-6mm}
    \caption{specBuf}
    \label{fig:specBuf}
\end{wrapfigure}
entry and a specBuf entry in both L1 and L2. Figure~\ref{fig:specBuf} shows the specBuf organization. For a given LQ entry in core(i)---$LQ[i,j]$---there is a corresponding specBuf entry
in L1 cache, $SB_{L1}[i,j]$, and L2 cache, $SB_{L2}[i,j]$.
We denote the specBuf of a core(i) in L1 and L2 as $SB_{L1}[i,*]$
and $SB_{L2}[i,*]$, respectively.
The {\em valid} bit indicates whether the entry is in use---only
the LQ entries for speculative loads have valid specBuf entries. 
The {\em ready} bit indicates whether the coherence transactions
related to the entry is in transit.
The {\em metadata} field keeps speculative access information, e.g.,
the number of accesses performed to the cache line while it is 
speculative, which is used to update the cache 
status if the line is merged later.
While we indicate {\em SpecData} field, it is only used to 
store the actual data of the cache line if it is not allocated in 
cache. 
Thus, there are not frequent data movements between 
specBuf and cache during merge.
Similarly, {\em Coh\_State} records the coherence state
of the line, and is only used when it does not 
exist in cache. Otherwise, the normal state field 
in each cache line is used to keep the state. 
The combined size of all $SB_{L1}[i,*]$ and $SB_{L2}[i,*]$
is $2 \times (\#\:of\:cores) \times (\#\:of\:LQ\:entries)$. 
The number of specBuf entries is 
the same as InvisiSpec~\cite{yan2018invisispec}.

The additional hardware structure associated with each $SB_{L2}[i,*]$ is a {\em counting bloom filter ($CBF_i$})~\cite{fan2000summary},
which approximately records the address set of cache lines
that present in each $SB_{L2}[i,*]$.
In CBF, addresses can be both inserted
and removed, thus maintaining a dynamic address set.
Using bloom filters, the membership check can be done very 
fast, it can generate false positives but never false negatives. 
The CBFs in L2's specBufs are used to maintain spec core information. 
Since all speculative loads
are recorded in specBuf of L2, this can be obtained by checking 
all $SB_{L2}[j,*]$, where $j \neq i$. 
The $CBF$s associated with each $SB_{L2}[j,*]$ can 
reduce the overheads by first performing
membership check of the line address with all $CBF_j$ ($j \neq i$),
and only performing search if the outcome is positive.
To prevent timing side-channels, 
we make the time to check the CBFs constant.

\subsection{Speculative Buffer Operations}
\label{spbuf_op}



The operations in L1/L2 are similar, we 
discuss base on L1. 
For a \rev{$Rd$}, if it misses in L1, 
no matter whether the line hits in specBuf,
a $GetS$ request is sent to L2 to bring the line to L1. 
For a \rev{$SpecRd$}, 
if it misses in both cache and specBuf, 
the line is brought to the specBuf, no cache 
block is allocated, and the state
is recorded in specBuf;
if it hits in cache but misses in specBuf,
the cache line is brought from cache to specBuf, 
the state is changed into a speculative state, 
and the cache line and specBuf entry
have the same state.
If it misses in cache but hits in specBuf,
there is no state change and the speculative load
gets data from specBuf. 

\section{Reversible Coherence Protocol}
\label{new_prot}

\subsection{Speculative States}
\label{sp_state}

Based on the insights discussed in Section~\ref{intro}, 
each state $X$ transitions to the corresponding
speculative state $XSpec$ on $SpecRd$ (in L1) or
$GetSpec$ (in L2). 
The speculative states can transition among 
each other triggered by non-speculative requests. 
Table~\ref{table:states} indicates the 
speculative states and 
the corresponding status in \projectname.

\begin{table}[h]
\begin{center}
\vspace{-0mm}
\footnotesize
\begin{tabular}{|p{0.7cm}||p{6.4cm}|}
\hline 
 States &  Global Status \\
\hline
 ISpec & No non-spec copy, one local spec copy  \\
\hline
  ESpec & One non-spec copy (E), one/more spec copies\\
\hline
  SSpec & Multiple non-spec copies, one/more spec copies\\
\hline
  MSpec & One non-spec copy (M), one or more spec copies \\
\hline
\end{tabular}
\end{center}
\vspace{-4mm}
\caption{Speculative States of \projectname}
\vspace{-2mm}
\label{table:states}
\end{table}

\subsection{Split-phase State Transition}
\label{split}

\begin{figure}[t]
    \centering
    \includegraphics[width=0.8\linewidth]{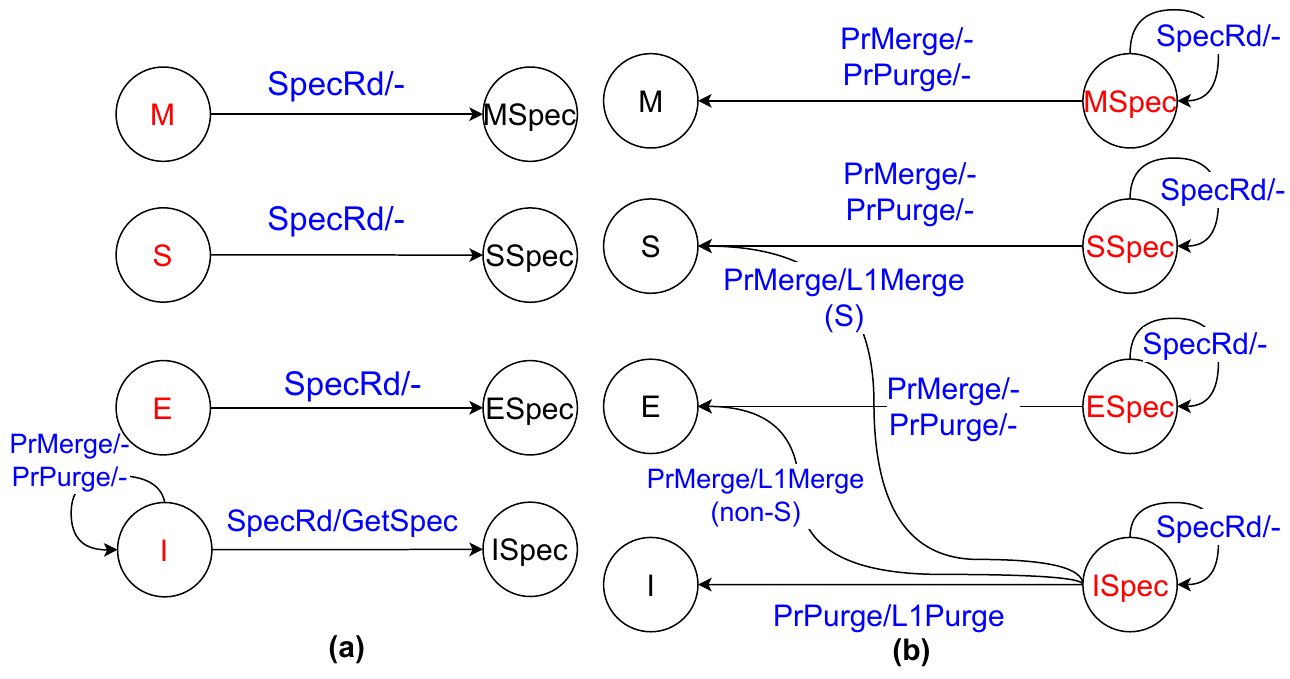} 
    \vspace{-2mm}
    \caption{Split-phase State Transition in L1}
    \vspace{-4mm}
    \label{split1}
\end{figure}

\rev{{\bf Insights} Figure~\ref{insight} (c) shows the insights with the 
simplistic scenario without conflicting accesses. 
When a $SpecRd$ is issued and the state of the 
requested cache line is $X$, it transitions to 
$XSpec$ ({\em speculative state}), where $X$ can be $M/E/S/I$.
After the transition, further $SpecRd$ will not 
trigger new transitions. 
Later, if $PrMerge$ is received, $XSpec$ transitions
to $X$ (if $X$ is $M/E/S$) or $S/E$ (if $X$ is $I$ depending
on the whether the line is shared).
If $PrPurge$ is received, $XSpec$ transitions back to $X$.
In MESI protocol, a load will directly
trigger $X \rightarrow X$ or $I \rightarrow S/E$ transitions
without going through $XSpec$.
With conflicting accesses, $XSpec$ will transition to other 
speculative states when a non-speculative or 
a forwarded speculative request accesses the line. }

{\bf \projectname specification}
Figure~\ref{split1} (a) shows the first phase
transition triggered by a speculative load. 
If the request misses
in L1, a $GetSpec$ request is sent to L2, and the 
transition $I \rightarrow ISpec$ is triggered by the 
response of $GetSpec$. 
For the other states, $SpecRd$ triggers the transition. 
In $I$, it is possible to receive $PrMerge$ or $PrPurge$
because the cache line is invalidated after $SpecRd$. 
In these cases, $PrMerge$ or $PrPurge$ are ignored. 

Figure~\ref{split1} (b) shows the second phase transition in L1 from speculative state back to non-speculative state
triggered by $PrMerge$ or $PrPurge$.
For $PrPurge$,  
$XSpec$, $X \in \{M,E,S\}$, will transition to 
$X$ because L1 still holds a non-speculative copy. 
For $PrMerge$ on $ISpec$, 
an $L1Merge$ is sent to L2 cache. 
If the line is shared (S), $ISpec$ will transition to $S$; 
otherwise (non-S), it will transition to $E$.
While the L1 cache is waiting for the response from L2,
it resolves the race condition by NACK-ing all requests
to the cache line. 

\begin{figure}[t]
    \centering
    \includegraphics[width=\linewidth]{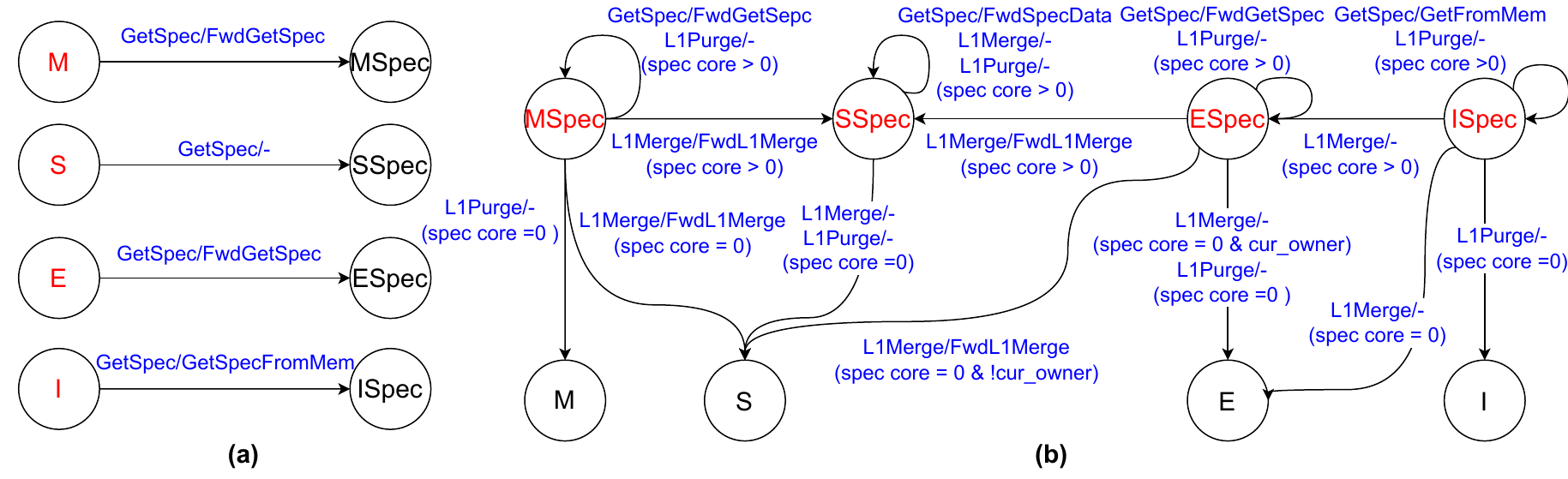} 
    \vspace{-6mm}
    \caption{Split-phase State Transition in L2}
    \vspace{-6mm}
    \label{split2}
\end{figure}

Figure~\ref{split2} shows the split-phase transitions
in L2. 
In Figure~\ref{split2} (a), 
$SpecRd$ missing in L1 triggers a $GetSpec$.
For $M/E$, $SpecRd$ is forwarded to the
current owner.
In Figure~\ref{split2} (b), 
The counter spec core is increased or decreased when a $GetSpec$
or an $L1Merge/L1Purge$ is received.
If spec core = 0 after an $L1Purge$, meaning
that there is no other speculative loads, each $XSpec$, $X \in \{M,E,S,I\}$,
will transition back to $X$; otherwise
they stay in the same state.

If spec core = 0 after an $L1Merge$, 
the speculative state transitions to a non-speculative state
as if a $GetS$ is received. 
For $MSpec$, it transitions to $S$ since there are more than one shared
copies, in addition, L2 forwards an $L1Merge$ to the current owner.
For $ESpec$, it can transition to $E$ or $S$.
The caveat is 
that the speculative and non-speculative copy
can be brought by the same processor.
It can be determined by 
comparing the current non-speculative owner
{\em cur\_owner} with the sender of $L1Merge$,
if they are the same, then $ESpec$ transitions
to $E$. Otherwise, it transitions to $S$, and
L2 also forwards an $L1Merge$ to cur\_owner.
Note that the cur\_owner information already 
exists in MESI to identify
the owner when a line is in $M/E$ in L2. 
In \projectname, cur\_owner is updated in two 
additional cases:
1) a $GetS$ is received in $ISpec$---a $Rd$ will get
the only non-speculative copy of the line in an L1; or 
2) $L1Merge$ is received in $ISpec$---a speculative copy
becomes the only non-speculative copy.

If spec core > 0 after an $L1Merge$,
there are still speculative copies, 
but there will be state transitions among speculative
states based on the number of 
non-speculative copies.
Specifically, $MSpec$ will transition to $SSpec$, because 
now we have more than one non-speculative copies plus some speculative copies.
For $SSpec$ and $ESpec$, 
they will both transition to $SSpec$, 
because there are more than one non-speculative copies
plus at least one speculative copy. 
For $ISpec$, it transitions to $ESpec$, because we have
one non-speculative copy (just merged) and at least one speculative 
copy.

\subsection{Non-interference of Speculative Load}
\label{non_inter}

\rev{{\bf Insights} When a $SpecRd$ misses in L1, $GetSpec$ is forwarded to the
current owner, in which the line can be in $X$ or $XSpec$, 
where $X$ is $M/E$. 
In \projectname, a $SpecRd$ does not cause state changes to 
remote caches since it will incur timing changes. 
Figure~\ref{insight} (d) shows this insight:
on receiving a $GetSpec$ on $X$ or $XSpec$ ($X=M/E$) the state
is not changed but data is forwarded; when later 
an $L1Merge$ is forwarded by L2, $X$ transitions to $S$
or $SSpec$ with data flushed to L2. 
In \projectname, $PrPurge$ is never forwarded (thus
no $L1Purge$ message) exactly because $GetSpec$ does not
cause state transition. 
In MESI protocol, $X$ will directly 
transition to $S$ on $GetS$.}

\begin{wrapfigure}[12]{r}{0.4\linewidth}
    \centering
    \includegraphics[width=\linewidth]{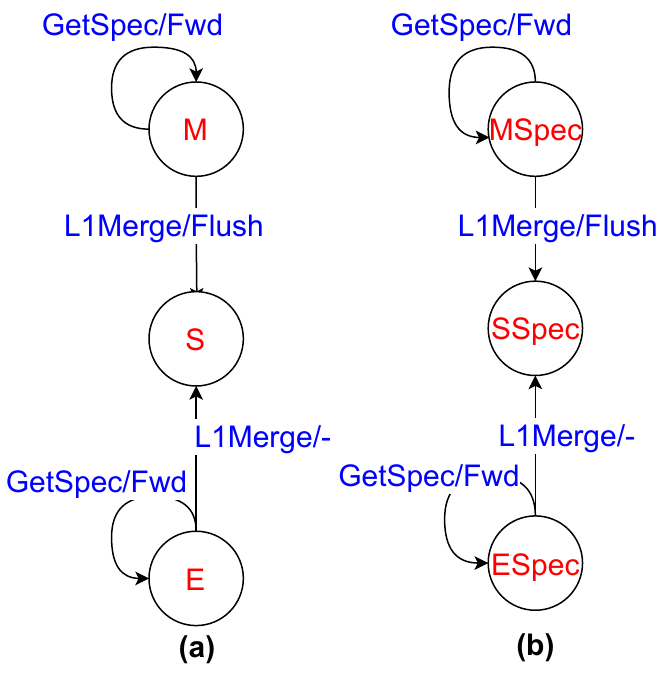} 
    \vspace{-6mm}
    \caption{Non-interference Property}
    \vspace{-6mm}
    \label{ni}
\end{wrapfigure}

{\bf \projectname specification} Figure~\ref{ni} shows the state transitions
that 
\setlength{\intextsep}{2pt}%
\setlength{\columnsep}{8pt}%
implement non-interference of speculative load,
of which the insight is illustrated
in Figure~\ref{insight} (d).
In Figure~\ref{ni} (a),
when L1 receives the forwarded $GetSpec$, the cache 
line can be in $M$/$E$.
The owner forwards the data but stays in the same state, otherwise
the speculative load will have effects.
When a remote $SpecRd$ is merged, the owner L1 cache 
will receive an $L1Merge$, which finalizes the transition 
(the second phase) to $S$.  
For $M$, the dirty line is flushed to L2 cache, which 
should be performed only when the speculative load becomes safe. 
In Figure~\ref{ni} (b),
the three speculative states' transitions
are the same as
the corresponding non-speculative states
that we described above.
The only difference is that, there is a pending
speculative load from the processor.

\subsection{Speculative State Transitions}
\label{spec_trans}

\begin{figure}[t]
    \centering
    \includegraphics[width=\linewidth]{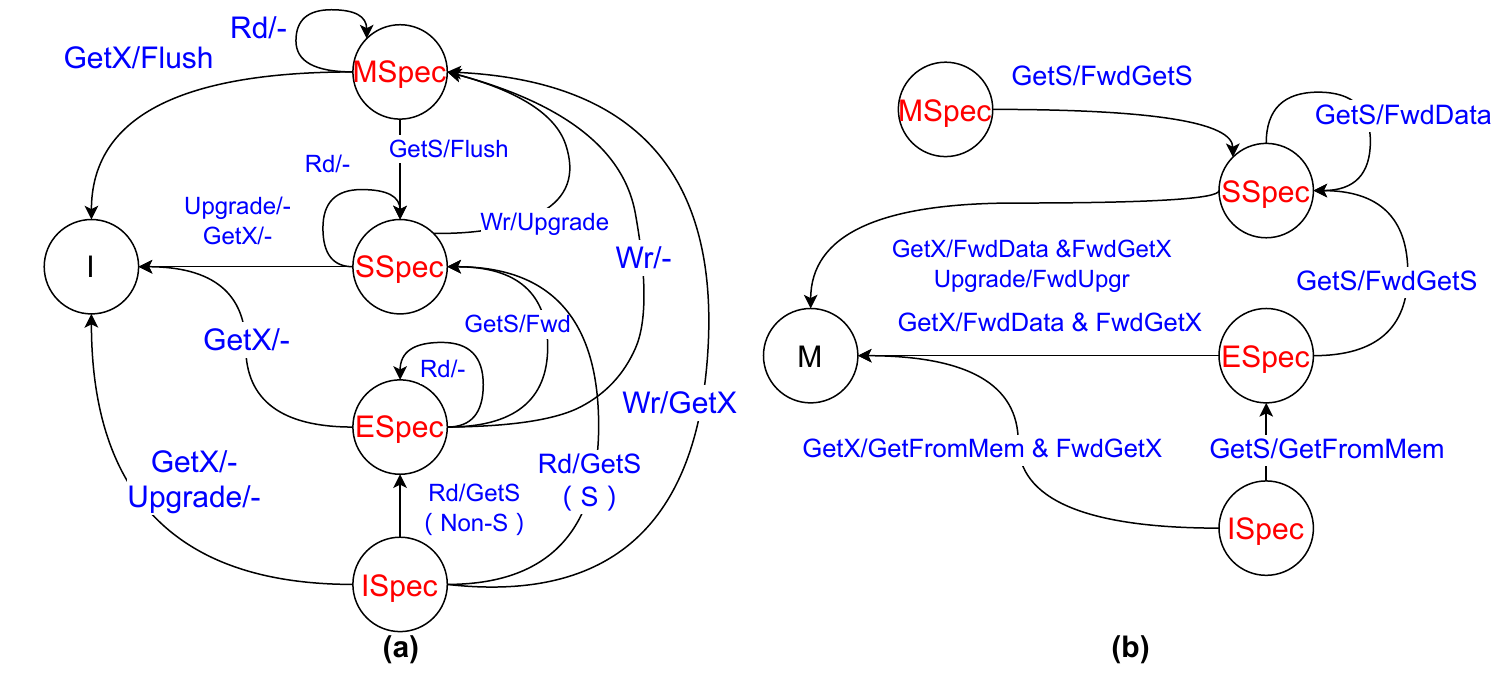} 
    \vspace{-6mm}
    \caption{Speculative State Transitions}
    \vspace{-6mm}
    \label{strans}
\end{figure}

\rev{{\bf Insights} \projectname needs to answer two questions:
(1) what happens when non-speculative requests access cache
lines in speculative states; and 
(2) what happens when $L1Merge$ are $L1Purge$ are
received in L2 (shared cache)?
Fortunately, the state transitions can be designed 
by mechanically extending the original MESI protocol. 
For (1), as indicated in Figure~\ref{insight} (e), 
a request will transition $XSpec$ to $YSpec$, if
the same request transitions $X$ to $Y$. 
Here $X,Y \neq I$ and they can be the same state. 
Moreover, we have $XSpec \rightarrow I$ corresponding to
each of $X \rightarrow I$ in MESI; 
and $ISpec \rightarrow XSpec$ for $I \rightarrow X$.
The protocol property is that the non-speculative requests  
trigger state transitions in the same manner as in
MESI but among corresponding speculative states. }

\rev{For question (2), when an $L1Purge$ or $L1Merge$ is received at L2, 
spec core is decreased; when a $getSpec$ is received, 
it is increased. 
As shown in Figure~\ref{insight} (f),
on receiving a $GetSpec$, $XSpec$ stays in the same state because a speculative load does not further change 
speculative state, similar to the case in 
Figure~\ref{insight} (c).
On receiving an $L1Purge$, 
if  spec core > 0 after the update,
$XSpec$ also stays in the same state because there
are other speculative copies;
if  spec core = 0, $XSpec$ transitions to
$X$---reversing back to the non-speculative state.
For $L1Merge$, the transition can be constructed 
on the transition triggered by $GetS$ in MESI.
For each $X \xrightarrow{GetS} Y$:
(1) if  spec core > 0, we have $XSpec \xrightarrow{L1Merge} YSpec$; 
(2) if  spec core = 0, we have 
$XSpec \xrightarrow{L1Merge} Y$.
The insight is: $LMerge$, indicating the {\em second} phase
of a speculative load, will trigger state transitions
in the same way as $GetS$ in MESI. }

{\bf \projectname specification} 
Figure~\ref{strans} shows the transitions among
speculative states due to non-speculative requests,
of which the insight is illustrated in Figure~\ref{insight} (d).
In Figure~\ref{strans} (a),
the key property is that the speculative states 
can transition among each other by non-speculative 
requests, and when a $PrMerge$ or $PrPurge$ is received, the state will return to 
the correct non-speculative states.
All speculative states transition to $I$ on receiving an invalidation 
($GetX$ or $Upgrade$), but processor still sends
$PrMerge$ or $PrPurge$ for the previous 
$SpecRd$
that caused the transition to a speculative state.
As shown in Figure~\ref{split1} (a), they should be ignored.
The speculative load may be re-issued 
depending on memory consistency model. More
details are discussed in Section~\ref{sec_analysis}.
In Figure~\ref{strans} (b) shows the similar 
specluative state transitions in L2.

\subsection{Non-Atomic Transactions}
\label{diss}

In \projectname, non-atomic
transactions in the original MESI protocol are handled
with the existing mechanisms---Nack-ing incoming requests 
while in transient states.
Among speculative states and requests, only
transitions related to 
$ISpec$ may lead to transient
period.
It happens when transitioning 
to $ISpec$ or from $ISpec$
to an $E/S$.
In L1, when a $SpecRd$ misses in 
L1, a $GetSpec$ request is sent
to L2 and when the response is received, the state transitions
to $ISpec$.
When L1 receives a \rev{$PrMerge$} at $ISpec$,
a request is sent to check whether
the line is shared or 
not, then $ISpec$ transitions to 
$E$ (non-S) or $S$ (S) accordingly.
These two cases can be handled
by keeping all requests during such 
transient period and only processing 
them when the state is stable.
We use 
a bit to indicate
that the cache line is in $ISpec$ but it is waiting
for the data ($I \rightarrow ISpec$) or non-S/S
information ($ISpec \rightarrow E/S$). 
When the bit is set, the incoming
requests are Nack-ed and processed later. 
The reason why \projectname is mostly not affected by non-atomic transactions is that,
the additional states and transitions are
introduced to handle only speculative {\em loads},
while most complications are due to writes. 

\rev{A relevant issue is the {\em ordering guarantee of messages} in the
on-chip network. For \projectname, the protocol state transitions
related to speculative states and messages, i.e., speculative load,
merge, and purge, do {\em not} assumes the ordering guarantee. 
For the normal state transitions in the original MESI protocol, 
we keep the assumption unchanged. In another word, \projectname does
not add additional message ordering requirement. 
\projectname is designed in a manner that one state can 
receive different potential messages in {\em any} orders, which 
will lead to different {\em but all correct} state transition paths. 
For example, consider $ESpec$ in an L1, it can ``concurrently''---{\em in any order}---receive a forwarded $GetSpec$ and a local $PrPurge$. In \projectname, processing them in either order is correct. 
If $PrPurge$ is processed first, the state transitions to $E$, and the cache will forward data as the response to the later $GetSpec$ but stays in $E$. If $GetSpec$ is processed first, the state is still $ESpec$, which will also transition to $E$ when $PrPurge$ is processed.}

\rev{Finally, we discuss the handling of {\em cache replacement}. 
The protocol state transitions described so far
assume that 
a cache block's state should eventually transition to a
non-speculative state. 
If a cache block in speculative state needs to be replaced, we simply 
allow that to happen. 
It is important to know that, after the replacement, 
the speculatively loaded data is still in the specBuf. 
Later, when the speculative load is merged or purged, 
the request misses in L1
but hits in specBuf, a $GetSpec$ request is sent to the L2.
The returned block from L2 has a bit indicating whether the block 
in L2 is shared (non-speculatively or speculatively). 
The block is re-inserted in L1 in $SSpec$ if it is shared in L2;
otherwise, the state is the same as the
state recorded in the specBuf entry. 
After that, the operations triggered by the merge or purge 
are applied and the block eventually transitions to 
a non-speculative state. 
In this case, the specBuf has the effect of 
enlarging L1 size, leading to potential speedups. 
For simplicity, we do not indicate such transition in the 
protocol specification. }

\begin{figure*}[!ht]
    \centering
    \includegraphics[width=\linewidth]{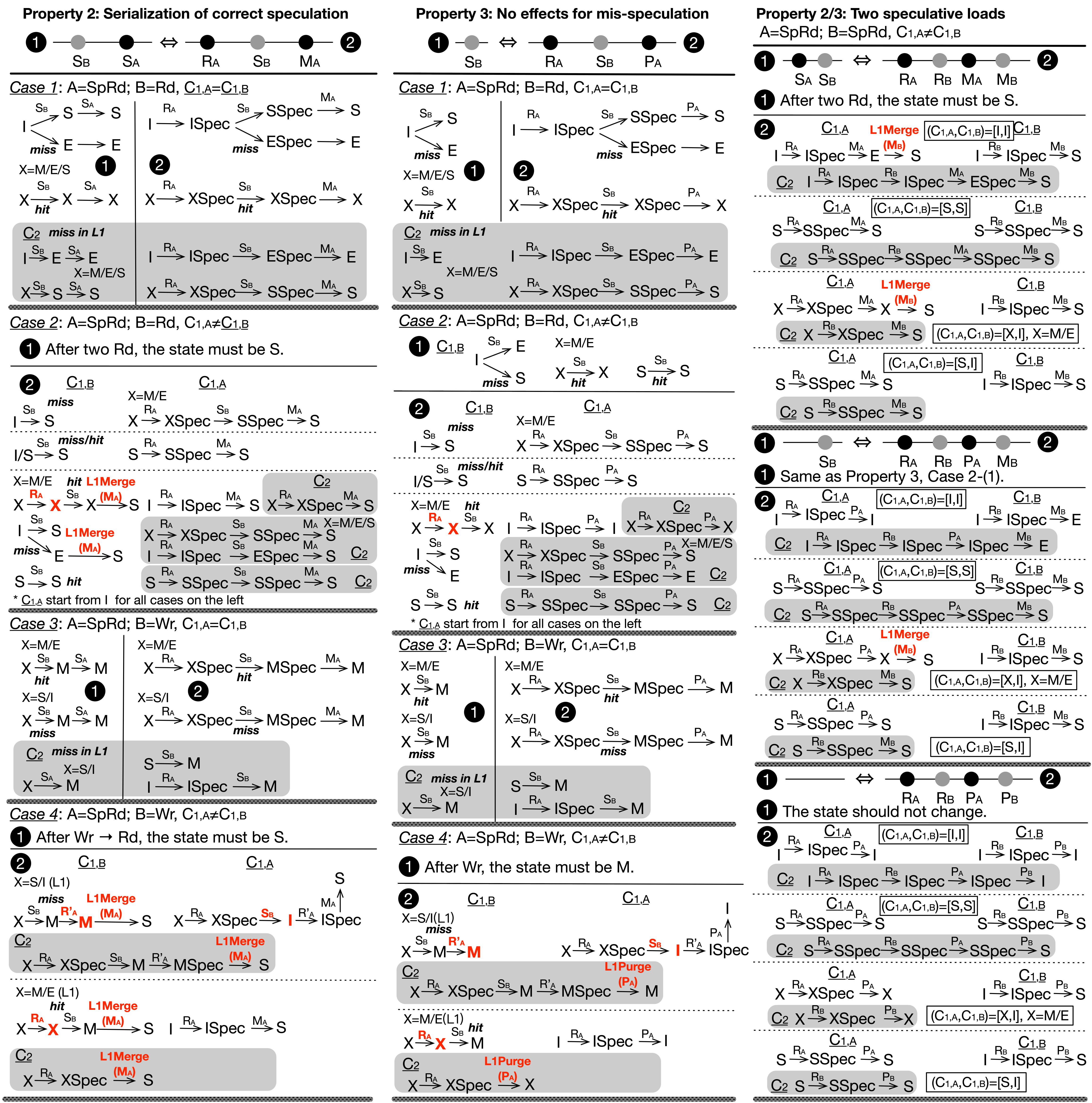}
    \caption{Proof of \projectname's Security Properties 2\&3}
    \label{fig:rcp_proof}
\end{figure*}

\subsection{Protocol Verification}
\label{veri}
We use Murphi~\cite{276232}
to verify our protocol and quantify the complexity of \projectname. 
We model the two-level MESI protocol from GEM5 implementation 
suite and make it as a baseline. 
To keep the number of explored states tractable, we use
a single address, two data
values with private L1 cache and a shared L2 with directory, similar to ~\cite{6113797}. With one address, we modeled
replacements as unconditional events that can be triggered at
any time. 
The verification explored 3,244,350 reachable states, which is 70\% more than MESI, in 1,186 seconds.
We verify that the protocol states  
transitions happen as designed.
The verification reaches all speculative states, 
which can always transition back to NS eventually.  
\section{Security Analysis}
\label{sec_analysis}

We show that \projectname satisfies
the three properties in Section~\ref{sec_prop} \rev{and 
the additional requirement that for the correct speculated load, 
the protocol exhibits the same behavior as a non-speculative load (overcapture requirement)}.
For a normal load/store $A$,
the serialization point ($S_A$)
is the point during its pending 
period that the access is serialized
in the global total order of 
accesses to this address. 
With atomic writes, $S_A$ is well-defined. 
A speculative load $A$ does not have a single serialization point, instead,
it has a {\em reach point ($R_A$)},
indicating the point that it reaches
L1/L2; and a {\em merge point ($M_A$)}
or {\em purge point ($P_A$)}, indicating
the point that it is merged or purged 
in L1/L2. 

{\bf Proof of Property 1 \rev{and the overcapture requirement}}:
Without overlapping, this property is 
ensured by design with the split-phase transition. 
The location property is ensured because a $SpecRd$ installs
the cache line to all cache components (specBuf if miss in a component)
as if it is a $Rd$. Depending on purge or merge, the specBuf entry
is simply removed or inserted into cache.

{\bf Proof of Property 2 ({\em SpecRd} overlaps with load/store)}:
We consider all possible behaviors
of a non-speculative access $B$ and 
a speculative load $A$, where
$R_A<S_B<M_A$.
We have four combinations: (a) $B$ is 
a load $Rd$ or a store $Wr$; and (b) they are
issued from the same or different 
L1 cache: $C_{1,A}=C_{1,B}$ or 
$C_{1,A} \neq C_{1,B}$.
Figure~\ref{fig:rcp_proof} (left) shows that for each case
and all possible initial states, 
the state and location of the cache 
line (\ding{183}) is the same as $B$ is serialized
before a non-speculative load $A$ (\ding{182}) to 
the same location; and the timing for the non-speculative
access does not change.

{\bf Proof of Property 3 ({\em SpecRd} overlaps with load/store)}:
In this case, $A$ is squashed 
so $R_A<S_B<P_A$, and the effects
should be equivalent to 
executing just $B$.
Figure~\ref{fig:rcp_proof} (middle)
shows such equivalence for all cases. 

{\bf Proof of Property 2 and 3 ({\em SpecRd} overlaps with another {\em SpecRd}}:
Between two
$SpecRd$ $A$ and $B$, Figure~\ref{fig:rcp_proof} (right)
shows the equivalence:
1) ($A$ and $B$ merge)=(two non-speculative loads serialized
according to the merge order);
2) ($A$ purges and $B$ merges)=(one non-speculative $B$); and
3) ($A$ purges and $B$ purges)=(no load).

{\bf Example: Explanation of Case 4 of Property 2. }
This case assumes 
Total-Store-Order (TSO) model, which prevents the reordering of loads.
When a write ($B$) is serialized before a non-speculative read ($A$) from a different core, the eventual state in $C_{1,A}$ and $C_{1,B}$ should be $S$ (\ding{182}).  
For \ding{183}, 
if the line is in $S$ or $I$ ($X=S/I$) 
in $C_{1,A}$ and $C_{1,B}$ (the case above dashed line),
it must be in $S$ or $I$ in $C_2$.
So $R_A$ will transition from $X$ to $XSpec$, $C_{1,B}$ is not affected.
When $B$ is serialized ($S_B$), the line in $C_{1,B}$ and $C_2$ both
transition to $M$. In $C_{1,A}$, it is invalidated. 
At this point, MCM matters: in TSO, if a load cache line is
invalidated before the load retires from the processor, it has 
to be ``replayed''. It is the ``Peekaboo'' problem discussed in 
~\cite{sorin2011primer}. The replay is needed because otherwise, the 
load might be reordered with an earlier load---prohibited by TSO.
Thus, the $SpecRd$ is re-issued to $C_{1,A}$, first reaching 
$ISpec$ ($C_2$ transitions from $M$ to $MSpec$), and finally 
reaching $S$ on merge in both $C_{1,A}$ and $C_2$.

If the cache line is already in an exclusive state
in the write $B$'s
cache $C_{1,B}$ ($X=M/E$ shown below dashed line), it must be $I$ in 
the $SpecRd$'s cache $C_{1,A}$. 
At $R_A$, $SpecRd$ misses in $C_{1,A}$ and $C_2$ forwards
it to $C_{1,B}$. \projectname provides the data but $M/E$ state
does not changes (emphasized in red).
At $C_{1,A}$, $R_A$ changes the state to $ISpec$.
After that, when the write is performed ($S_B$), it can still 
hit in $C_{1,B}$---not affected by $SpecRd$.
When $SpecRd$ is merged, the state in both $C_{1,A}$
and $C_2$ transition to $S$.
In both cases ($X=S/I$ or $X=M/E$), 
final state and location of the cache line are
the same as \ding{182}.

{\bf \projectname and memory consistency models (MCMs). }
We also use the earlier case to show 
\projectname works with release consistency (RC).
In RC, since loads to different locations can be reordered, 
even if a load cache line is invalidated before
it retires from the processor, the load do not need to be replayed. 
In this case, for \ding{183} the final state in $C_{1,B}$, $C_{1,A}$, and $C_2$
are $M$, $I$, $M$, respectively. 
However, it is not inconsistent with \ding{182}, because, without
the load replay, the read $B$ is essentially serialized {\em before}
the write $A$ ($Rd \rightarrow Wr$), and the eventual states
are the same. In RC, without load replay, it is
not possible to have $Wr \rightarrow Rd$.

{\bf Non-atomic Writes. }
\projectname works with processors (e.g., IBM PowerPC) that do not enforce write atomicity, i.e., stores
are visible to different cores at different time. 
\projectname supports invisible 
speculative load, which only interacts with stores when it is 
invalidated. 
Even with non-atomic writes, the write performance w.r.t. 
{\em individual load} is always well-defined. 

Since coherence and MCM are interlinked~\cite{manerkar2015ccicheck}, 
\projectname needs to be sufficient to support correct synchronization.
Consider the WRC litmus test~\cite{trippel2017tricheck}:
[C0: St x 1; C1: Ld x 1; St y 1; C2: Ld y 1; Ld x 0].
With non-atomic writes, such unintuitive behavior is possible since
the write from C0 is performed w.r.t. C1 before C2. 
To achieve the desired behavior, e.g., ensuring the load in C2 returns 1,
fences need to be {\em cumulative}, which enforce ordering between
instructions across cores. 
In this case, the cumulative fence should be inserted between
the two instructions in C1, and forces the write to perform only 
after the read from $x$ is globally performed. 
Implementing such semantics requires C0 to collect all 
invalidation acknowledgements (from both C1 and C2) and then
notify C1 so that it can perform the write to $y$. 
In \projectname, as in MESI, 
an acknowledgement is sent after the cache line
is invalidated. Moreover, invalidation does not affect speculative state
change: the line always transitions to $I$, no matter it was in NSS or SS.
Thus, we believe \projectname is not affected by non-atomic writes and
can correctly implement the necessary synchronizations.

{\bf Whole-program Proof Sketch}: All proofs consider
two instructions, but the security
property of the whole program
can be inductively obtained.
Specifically, we can consider the two-instruction 
case as the initial case in mathematical induction, then we
can assume that these properties hold with a sequence less than $n$
instructions, we need to prove they are true with $(n+1)$
instructions. The crux of the argument is that, the first 
$(n-1)$ instructions satisfy the properties (inductive assumption), 
and they will produce a state and location setting 
denoted by $\mathcal{S}_{n-1}$ and $\mathcal{L}_{n-1}$ satisfying 
the properties. Then the last two instructions $n$ and $(n+1)$
execute from $\mathcal{S}_{n-1}$ and $\mathcal{L}_{n-1}$, 
preserving the properties based on our proof.

{\bf Security of SpecBuf}. We show that specBuf cannot be used to create new channels. 
Referring to Section~\ref{specBuf}, there is a {\em one-to-one mapping} between a core's LQ 
entry, $LQ[i,j]$, 
to specBuf entries in L1 $SB_{L1}[i,j]$ and L2 $SB_{L2}[i,j]$. 
There is no fully associative or set-associative
hardware structures, thus it is 
{\em not} vulnerable to cache-based side-channel attacks such as Prime+Probe. 
In fact, the specBuf organization is exactly the same as InvisiSpec---with the same number of entries---expect the bloom filters in L2's specBuf and some 
counters for each entry. 
In \projectname, we make 
the time to check the bloom filter
in L2 constant, so the present/absence information
in the specBuf cannot be revealed through 
the time for the check.

{\bf Security of SpecBuf}. We show that specBuf cannot be used to create new channels. 
Referring to Section~\ref{specbuffer}, there is a {\em one-to-one mapping} between a core's LQ 
entry, $LQ[i,j]$, 
to specBuffer entries in L1 $SB_{L1}[i,j]$ and L2 $SB_{L2}[i,j]$. 
There is no fully associative or set-associative
hardware structures, thus it is 
{\em not} vulnerable to cache-based side-channel attacks such as Prime+Probe. 
In fact, the specBuffer organization is exactly the same as InvisiSpec---with the same number of entries---expect the bloom filters in L2's specBuf and some 
counters for each entry. 
\red{In \projectname, we make 
the time to check the bloom filter
in L2 constant, so the present/absence information
in the specBuf cannot be revealed through 
the time for the check.}

{\bf Case Study}. In Figure~\ref{fig:exist_problems} (f),
consider the case that the probed address from $P_2$ and $P_3$ is the same
as $SpecRd$ (correct guess). 
The key observation is that while $P_1$'s $SpecRd$ causes
state transitions at L1 and L2, when $P_2$'s $Rd$ is performed, it
behaves as if $SpecRd$ does not exist, both in terms of state
($I \rightarrow E$) and latency (miss in both L1 and L2). 
When $SpecRd$ is squashed, \projectname reverses the state in L1 and L2.
Later $P_3$'s $Rd$ get the line forwarded by $P_2$'s cache, exactly the same 
as in Figure~\ref{fig:exist_problems} (a) when the guess is wrong. 
\projectname prevents the 
attack and the final states and locations of the cache line
are the same as if only $Rd$ from $P_2$ and $P_3$ are performed. 

Figure~\ref{fig:exist_problems} (g) shows the execution when 
$P_1$'s $SpecRd$ is committed after $P_3$'s access.
When $SpecRd$ in $P_1$ commits, $C_{1,1}$ checks
with $C_2$ and finds that it should transition to $S$, then
sends an $L1Merge$ to $C_2$. There is only one speculative
copy, so  spec core=1, after the merge, it becomes 0. 
Thus, the state transitions
from $SSpec$ to $S$ with three non-speculative copies.
In Figure~\ref{fig:exist_problems} (h),
when $GetSpec$
is received on $M$ state in $C_{1,1}$, it can forward data
but {\em does not transition to $S$}.
This mechanism does not stall
$SpecRd$ on remote E/M.
When $SpecRd$ is committed, an $L1Merge$ is forwarded 
to $C_{1,1}$, and $M$ will transition to $S$ and flush
the data to $C_2$.

\section{Evaluation}

\begin{figure*}[h]
    \centering
    \vspace{-0.0cm}
    \includegraphics[width=\linewidth]{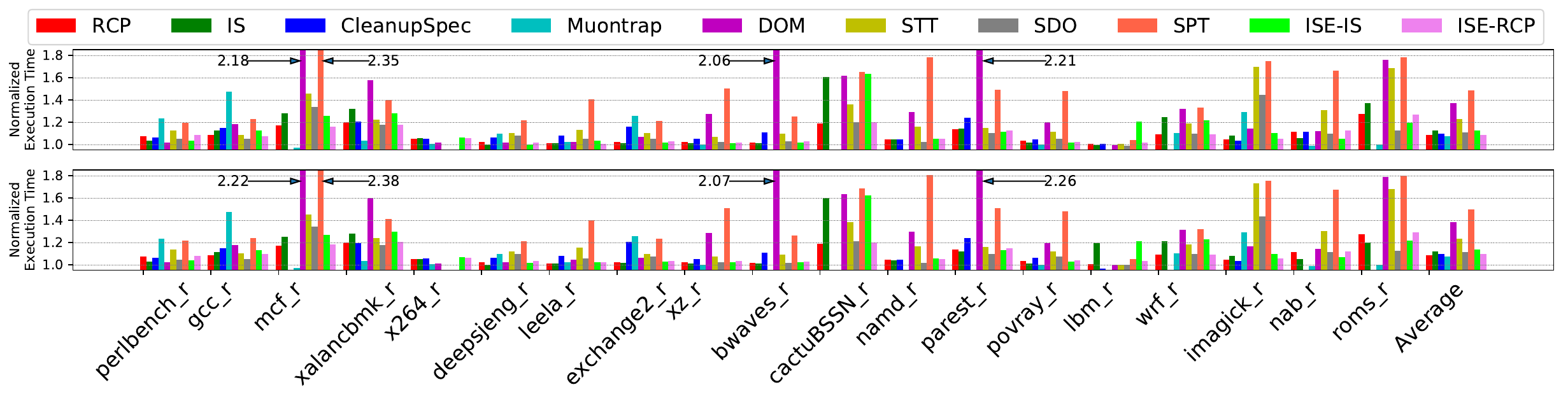}
    \vspace{-0.8cm}
    \caption{\rev{SPEC2017: Performance Overheads under TSO (upper) and RC (bottom)}}
    \label{fig:spec17}
\end{figure*}

\subsection{Environment Setup}
\begin{table}[t]

\footnotesize
\begin{tabular}{||p{1.6cm}||p{5.8cm}||}
\hline
\hline
Architecture & 1 core (SPEC) or 4 cores (PARSEC) at 2.0GHz \\
\hline
Core & 8-issue, OOO, no SMT, 32 LSQ entries, 192 ROB, Tournament Branch Pred., 4096 BTB, 16 RAS   \\
\hline
Private L1-I &  32KB, 64B line, 4-way, 1 cycle RT lat., 1 port  \\
\hline
Private L1-D & 64KB, 64B line, 8-way, 1 cycle RT lat., 3 ports   \\
\hline
Shared L2 &  2MB bank, 64B line, 16-way, 8 cycles RT local
latency, 16 cycles RT remote latency (max)   \\
\hline
Network & 4x2 mesh, 128b link width, 1 cycle per hop \\
\hline
Coherence & \projectname and MESI\\
\hline
DRAM & RT latency: 50 ns after L2 \\
\hline
\hline
\end{tabular}
\caption{Architecture Configurations}
\vspace{-6mm}
\label{table:Configurations}
\end{table}

We implemented \projectname and various
\ise in GEM5~\cite{binkert2011gem5}. 
To evaluate ISLE, we compare \projectname with 
of InvisiSpec (corrected), CleanupSpec using their public open source codes, \rev{MuonTrap}.
For InvisiSpec, we evaluate Futuristic model. For CleanupSpec, we evaluate the scheme for Cleanup\_FOR\_L1L2.
To evaluate \ise, we use \projectname and InvisiSpec
together with processor mechanisms described in Section~\ref{model}, they are
named {\em \ise-\projectname} and {\em \ise-IS}.
We compare the performance of \ise with 
STT, SDO and SPT under the same framework. 
The simulator configuration is shown in Table~\ref{table:Configurations}.
\rev{For all experiments, we use the x86 ISA, while some previous
papers used ARM ISA, this may lead to different performance overhead
than reported in these papers.}

We use SPEC CPU2017~\cite{bucek2018spec} for single-core evaluation, and PARSEC 2.1~\cite{bienia2008parsec} for multi-core evaluation.
For SPEC CPU2017, we run 19 workloads in intrate and fprate suits with reference input size.
We forward execution by 10 billion instructions and simulate 1 billion instructions. For PARSEC, we run 9 of the multi-threaded (4 cores) workload with the simmedium input size. 
We test two different memory consistency model, RC and TSO for SPEC2017;
and only show the results of TSO for PARSEC due to the space limitation.


\subsection{Performance Analysis}

\begin{figure}[h]
    \centering
    \includegraphics[width=\linewidth]{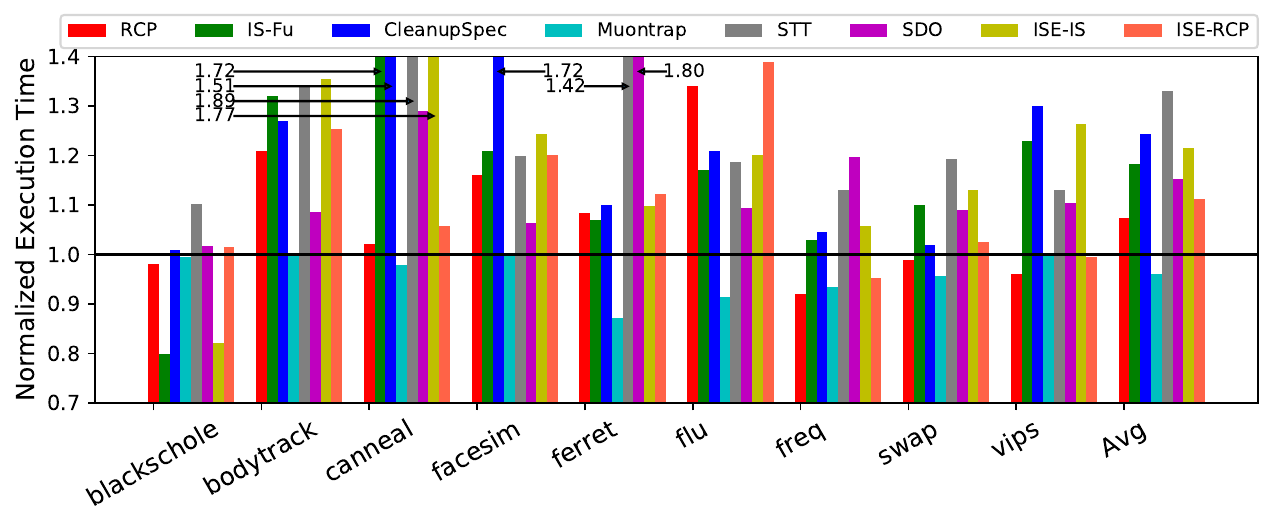}
    \caption{\rev{PARSEC: Performance Overheads under TSO}}
    \label{fig:PARSEC}
\end{figure}

Figure~\ref{fig:spec17} shows single core performance overheads
of (1) ISLE: \projectname, InvisiSpec, CleanupSpec, \rev{Muontrap} and DOM; 
(2) \ise: {\em \ise-\projectname} and {\em \ise-IS};  and 
(3) STT, SDO, SPT on SPEC2017 under TSO and RC, normalized to a non-secure baseline. We observe that both TSO and RC share a similar trend of performance overheads.
Overall, TSO incurs slightly lower overheads. 
Some of the benchmarks failed to run for CleanupSpec (e.g., mcf\_r and cactusBSSN\_r) and STT (x264\_r), we leave a blank bar for these cases. 
Figure~\ref{fig:PARSEC} shows the multicore performance overheads under TSO. 
The results under RC show similar trends. 

{\bf ISLE Overheads.}
Under TSO, \projectname incurs an average slowdown of 
7.7\%, while CleanupSpec,\rev{Muontrap} and InvisiSpec incur average slowdowns 
of 9.3\% \rev{, 7.6\%} and 12.6\%, respectively.
Under RC, the trends are similar, the slowdown of 
\projectname, CleanupSpec, \rev{Muontrap} and InvisiSpec are 
8.3\%, 9.4\%, \rev{7.7\%} and 12\%, respectively. 
For PARSEC, \projectname, CleanupSpec and InvisiSpec incurs 7.4\%, 24.2\% and 18.3\% slowdown on average \rev{, while Muontrap achieve 4\% speedup. This speedup is due to the filter L0 cache on top of the two level cache hierarchy.}
The DOM incurs much higher slowdown about 37.1\% and 38.4\% under TSO and RC. 
It can be reduced with further optimizations~\cite{tran2020clearing}. Due to the larger gap,
we exclude DOM in the comparison.

\projectname incurs lower overheads because it neither issues the second load nor stalls the processor during squash. 
The overheads of \projectname are mainly caused by the 
increased traffic due to merge/purge operations. 
Depending on mis-speculation rates, 
\projectname is not always better than InvisiSpec and CleanupSpec.
Overall, CleanupSpec incurs lower overheads with lower mis-prediction rate.  
\projectname is less sensitive to mis-prediction rate since
the merge/purge operations can be performed concurrently with 
execution. 
For example, lbm\_r has an extremely low mis-prediction rate (0.36\%), \projectname has negligible slowdown and CleanupSpec even has a decent speedup.
By design, CleanupSpec should never lead to speedups. 
A probable reason is that some codes of InvisiSpec were not completely removed (CleanupSpec is 
modified from InvisiSpec). 
Overall, for both RC and TSO, \projectname incurs the lowest overheads.

{\bf \ise Overheads.}
Under TSO, {\em \ise-\projectname} and {\em \ise-IS} incur on average 
(8.4\%,11\%) and 
(12.7\%,18.7\%) on (SPEC2017,
PARSEC), respectively. 
Under RC, {\em \ise-\projectname} and {\em \ise-IS} incurs on average 9.5\% and 13.5\%, which is slightly higher than TSO. 
The additional overheads to ensure {\bf \ise-P1} and 
{\bf \ise-P2} are only (0.7\%,3.6\%) based on \projectname, and 
(0.1\%,0.4\%) based on InvisiSpec under (TSO and RC). 



{\bf \ise and STT/SDO/SPT.}
Under TSO,
STT and SDO incurs (22.5\%, 32.2\%) and (10.9\%,15.1\%) overheads on (SPEC2017,PARSEC), respectively.
{\em \ise-\projectname} always incurs lower overheads than both STT and SDO
{\em even it provides slightly stronger security protection}. 
Under RC, STT and SDO incurs 23.2\% and 11.3\% overheads, respectively. 
SPT incurs about 48.5\% and 49.6\%---considerably higher than {\em \ise-\projectname}---to ensure the same security property. 


{\bf ``Speedup'' Analysis.}
From Figure~\ref{fig:PARSEC}, some benchmarks, 
e.g., blackschole under InvisiSpec and freq under \projectname,
runs faster with ISLE. It is due to the positive ``side effects''
of specBuf. 
If a $SpecRd$ issued to L1 cache triggers replacement of a cacheline in a non-secure processor, the specBuf will temporally 
keep the cache line and the replacement is avoided. 
As a result, it is possible that an access misses in L1 for non-secure 
processor while a speculative load (later becomes safe) hits in specBuf.
\projectname can fully enjoy this benefit given that merge/purge are
not in the critical path. 
For InvisiSpec, the negative effects of redo access may overshadow
the benefit, e.g., incurring overheads in freq and vips.


\vspace{-1mm}
\subsection{Coherence Traffic Analysis}

\setlength{\intextsep}{8pt}%
\setlength{\columnsep}{0pt}%
\begin{figure}[h]
    \centering
    \includegraphics[width=\linewidth]{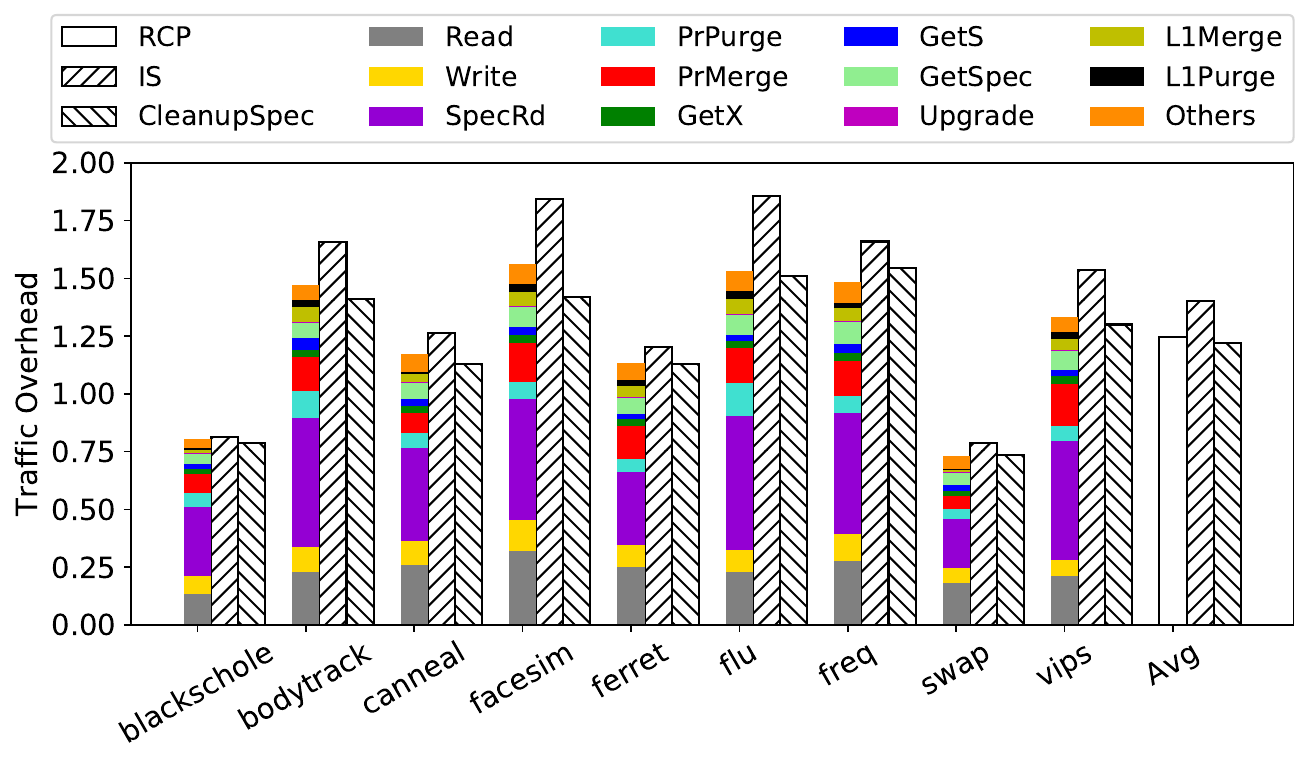}
    \caption{\rev{Traffic Overheads}}
    \label{fig:PARSEC_Traffic}
\end{figure}
Figure~\ref{fig:PARSEC_Traffic} compares 
the traffic overheads of
\projectname, InvisiSpec and CleanupSpec
under TSO,
normalized to the baseline MESI protocol. We measure the traffic overheads by counting the total number of bytes transferred among the cache system and between cache and the main memory. For \projectname, we also count the bytes for each type of the message and show their distribution in Figure~\ref{fig:PARSEC_Traffic}. 
\projectname 
incurs on average \rev{26\%} traffic overheads, while InvisiSpec and CleanupSpec incur 40\% and \rev{20\%}, respectively. 
Coherence traffic overheads of \projectname
is mainly attributed to 
$PrMerge$, $PrPurge$, and $L1Merge$ messages.
\rev{Compared to InvisiSpec, the traffic is decreased because we do not have any traffic caused by the validation and exposure operations. 
For a correct path, the merge of data does not introduce redundant data movement.}
Moreover, 
the merge and purge are performed  
{\em concurrently with processor execution}.

\vspace{-1mm}
\section{Conclusion}
\vspace{-1mm}

This paper proposes \projectname, a new reversible 
coherence protocol that ensures invisible 
speculative load execution (ISLE) with low overheads. 
\projectname can be combined with 
processor mechanisms that eliminate the 
effects of speculative instructions on other 
instructions to achieve low-overhead
invisible speculative execution (ISE).
\projectname is designed by systemtically
extending the existing 
coherence protocol to incorporate speculative 
loads and states. 
The results show that \projectname based ISE
incurs much lower overheads than STT/SPT while
providing similar or stronger protection.

\bibliographystyle{IEEEtranS}
\bibliography{refs}

\begin{thebibliography}{10}
\providecommand{\url}[1]{#1}
\csname url@samestyle\endcsname
\providecommand{\newblock}{\relax}
\providecommand{\bibinfo}[2]{#2}
\providecommand{\BIBentrySTDinterwordspacing}{\spaceskip=0pt\relax}
\providecommand{\BIBentryALTinterwordstretchfactor}{4}
\providecommand{\BIBentryALTinterwordspacing}{\spaceskip=\fontdimen2\font plus
\BIBentryALTinterwordstretchfactor\fontdimen3\font minus
  \fontdimen4\font\relax}
\providecommand{\BIBforeignlanguage}[2]{{%
\expandafter\ifx\csname l@#1\endcsname\relax
\typeout{** WARNING: IEEEtranS.bst: No hyphenation pattern has been}%
\typeout{** loaded for the language `#1'. Using the pattern for}%
\typeout{** the default language instead.}%
\else
\language=\csname l@#1\endcsname
\fi
#2}}
\providecommand{\BIBdecl}{\relax}
\BIBdecl

\bibitem{ainsworth2021ghostminion}
S.~Ainsworth, ``Ghostminion: A strictness-ordered cache system for spectre
  mitigation,'' in \emph{MICRO-54: 54th Annual IEEE/ACM International Symposium
  on Microarchitecture}, 2021, pp. 592--606.

\bibitem{ainsworth2020muontrap}
S.~Ainsworth and T.~M. Jones, ``Muontrap: Preventing cross-domain spectre-like
  attacks by capturing speculative state,'' in \emph{2020 ACM/IEEE 47th Annual
  International Symposium on Computer Architecture (ISCA)}.\hskip 1em plus
  0.5em minus 0.4em\relax IEEE, 2020, pp. 132--144.

\bibitem{behnia2021speculative}
M.~Behnia, P.~Sahu, R.~Paccagnella, J.~Yu, Z.~N. Zhao, X.~Zou,
  T.~Unterluggauer, J.~Torrellas, C.~Rozas, A.~Morrison \emph{et~al.},
  ``Speculative interference attacks: Breaking invisible speculation schemes,''
  in \emph{Proceedings of the 26th ACM International Conference on
  Architectural Support for Programming Languages and Operating Systems}, 2021,
  pp. 1046--1060.

\bibitem{bienia2008parsec}
C.~Bienia, S.~Kumar, J.~P. Singh, and K.~Li, ``The parsec benchmark suite:
  Characterization and architectural implications,'' in \emph{Proceedings of
  the 17th international conference on Parallel architectures and compilation
  techniques}.\hskip 1em plus 0.5em minus 0.4em\relax ACM, 2008, pp. 72--81.

\bibitem{binkert2011gem5}
N.~Binkert, B.~Beckmann, G.~Black, S.~K. Reinhardt, A.~Saidi, A.~Basu,
  J.~Hestness, D.~R. Hower, T.~Krishna, S.~Sardashti, R.~Sen, K.~Sewell,
  M.~Shoaib, N.~Vaish, M.~D. Hill, and D.~A. Wood, ``The gem5 simulator,''
  \emph{ACM SIGARCH Computer Architecture News}, vol.~39, no.~2, pp. 1--7,
  2011.

\bibitem{bucek2018spec}
J.~Bucek, K.-D. Lange, and J.~v.~Kistowski, ``Spec cpu2017: Next-generation
  compute benchmark,'' in \emph{Companion of the 2018 ACM/SPEC International
  Conference on Performance Engineering}, 2018, pp. 41--42.

\bibitem{6113797}
B.~{Choi}, R.~{Komuravelli}, H.~{Sung}, R.~{Smolinski}, N.~{Honarmand}, S.~V.
  {Adve}, V.~S. {Adve}, N.~P. {Carter}, and C.~{Chou}, ``Denovo: Rethinking the
  memory hierarchy for disciplined parallelism,'' in \emph{2011 International
  Conference on Parallel Architectures and Compilation Techniques}, 2011, pp.
  155--166.

\bibitem{choudhary2021speculative}
R.~Choudhary, J.~Yu, C.~Fletcher, and A.~Morrison, ``Speculative privacy
  tracking (spt): Leaking information from speculative execution without
  compromising privacy,'' in \emph{MICRO-54: 54th Annual IEEE/ACM International
  Symposium on Microarchitecture}, 2021, pp. 607--622.

\bibitem{deng2021leaky}
S.~Deng, B.~Huang, and J.~Szefer, ``Leaky frontends: Micro-op cache and
  processor frontend vulnerabilities,'' \emph{arXiv preprint arXiv:2105.12224},
  2021.

\bibitem{276232}
D.~L. {Dill}, A.~J. {Drexler}, A.~J. {Hu}, and C.~H. {Yang}, ``Protocol
  verification as a hardware design aid,'' in \emph{Proceedings 1992 IEEE
  International Conference on Computer Design: VLSI in Computers Processors},
  1992, pp. 522--525.

\bibitem{fan2000summary}
L.~Fan, P.~Cao, J.~Almeida, and A.~Z. Broder, ``Summary cache: a scalable
  wide-area web cache sharing protocol,'' \emph{IEEE/ACM transactions on
  networking}, vol.~8, no.~3, pp. 281--293, 2000.

\bibitem{fustos2020spectrerewind}
J.~Fustos, M.~Bechtel, and H.~Yun, ``Spectrerewind: Leaking secrets to past
  instructions,'' in \emph{Proceedings of the 4th ACM Workshop on Attacks and
  Solutions in Hardware Security}, 2020, pp. 117--126.

\bibitem{khasawneh2019safespec}
K.~N. Khasawneh, E.~M. Koruyeh, C.~Song, D.~Evtyushkin, D.~Ponomarev, and
  N.~Abu-Ghazaleh, ``Safespec: Banishing the spectre of a meltdown with
  leakage-free speculation,'' in \emph{2019 56th ACM/IEEE Design Automation
  Conference (DAC)}.\hskip 1em plus 0.5em minus 0.4em\relax IEEE, 2019, pp.
  1--6.

\bibitem{kim2021uc}
J.~Kim, H.~Jang, H.~Lee, S.~Lee, and J.~Kim, ``Uc-check: Characterizing
  micro-operation caches in x86 processors and implications in security and
  performance,'' in \emph{MICRO-54: 54th Annual IEEE/ACM International
  Symposium on Microarchitecture}, 2021, pp. 550--564.

\bibitem{kocher2019spectre}
P.~Kocher, J.~Horn, A.~Fogh, D.~Genkin, D.~Gruss, W.~Haas, M.~Hamburg, M.~Lipp,
  S.~Mangard, T.~Prescher, M.~Schwarz, and Y.~Yarom, ``Spectre attacks:
  Exploiting speculative execution,'' in \emph{2019 IEEE Symposium on Security
  and Privacy (SP)}.\hskip 1em plus 0.5em minus 0.4em\relax IEEE, 2019, pp.
  1--19.

\bibitem{koruyeh2018spectre}
E.~M. Koruyeh, K.~N. Khasawneh, C.~Song, and N.~Abu-Ghazaleh, ``Spectre
  returns! speculation attacks using the return stack buffer,'' in \emph{12th
  $\{$USENIX$\}$ Workshop on Offensive Technologies ($\{$WOOT$\}$ 18)}, 2018.

\bibitem{koruyeh2020speccfi}
E.~M. Koruyeh, S.~H.~A. Shirazi, K.~N. Khasawneh, C.~Song, and N.~Abu-Ghazaleh,
  ``Speccfi: Mitigating spectre attacks using cfi informed speculation,'' in
  \emph{2020 IEEE Symposium on Security and Privacy (SP)}.\hskip 1em plus 0.5em
  minus 0.4em\relax IEEE, 2020, pp. 39--53.

\bibitem{mengming_li_2021_5771649}
\BIBentryALTinterwordspacing
M.~Li, C.~Miao, Y.~Yang, and K.~Bu, ``{unXpec: Breaking Undo-based Safe
  Speculation - Artifact Evaluation for HPCA 2022},'' Dec. 2021. [Online].
  Available: \url{https://doi.org/10.5281/zenodo.5771649}
\BIBentrySTDinterwordspacing

\bibitem{lipp2018meltdown}
M.~Lipp, M.~Schwarz, D.~Gruss, T.~Prescher, W.~Haas, A.~Fogh, J.~Horn,
  S.~Mangard, P.~Kocher, D.~Genkin \emph{et~al.}, ``Meltdown: Reading kernel
  memory from user space,'' in \emph{27th $\{$USENIX$\}$ Security Symposium
  ($\{$USENIX$\}$ Security 18)}, 2018, pp. 973--990.

\bibitem{loughlin2021dolma}
K.~Loughlin, I.~Neal, J.~Ma, E.~Tsai, O.~Weisse, S.~Narayanasamy, and
  B.~Kasikci, ``$\{$DOLMA$\}$: Securing speculation with the principle of
  transient non-observability,'' in \emph{30th $\{$USENIX$\}$ Security
  Symposium ($\{$USENIX$\}$ Security 21)}, 2021.

\bibitem{manerkar2015ccicheck}
Y.~A. Manerkar, D.~Lustig, M.~Pellauer, and M.~Martonosi, ``Ccicheck: Using
  $\mu$hb graphs to verify the coherence-consistency interface,'' in \emph{2015
  48th Annual IEEE/ACM International Symposium on Microarchitecture
  (MICRO)}.\hskip 1em plus 0.5em minus 0.4em\relax IEEE, 2015, pp. 26--37.

\bibitem{ren2021see}
X.~Ren, L.~Moody, M.~Taram, M.~Jordan, D.~M. Tullsen, and A.~Venkat, ``I see
  dead $\mu$ops: Leaking secrets via intel/amd micro-op caches,'' in \emph{2021
  ACM/IEEE 48th Annual International Symposium on Computer Architecture
  (ISCA)}.\hskip 1em plus 0.5em minus 0.4em\relax IEEE, 2021, pp. 361--374.

\bibitem{saileshwar2019cleanupspec}
G.~Saileshwar and M.~K. Qureshi, ``Cleanupspec: An undo approach to safe
  speculation,'' in \emph{Proceedings of the 52nd Annual IEEE/ACM International
  Symposium on Microarchitecture}.\hskip 1em plus 0.5em minus 0.4em\relax ACM,
  2019, pp. 73--86.

\bibitem{sakalis2019efficient}
C.~Sakalis, S.~Kaxiras, A.~Ros, A.~Jimborean, and M.~Sj{\"a}lander, ``Efficient
  invisible speculative execution through selective delay and value
  prediction,'' in \emph{Proceedings of the 46th International Symposium on
  Computer Architecture}.\hskip 1em plus 0.5em minus 0.4em\relax ACM, 2019, pp.
  723--735.

\bibitem{sorin2011primer}
D.~J. Sorin, M.~D. Hill, and D.~A. Wood, ``A primer on memory consistency and
  cache coherence,'' \emph{Synthesis lectures on computer architecture},
  vol.~6, no.~3, pp. 1--212, 2011.

\bibitem{tran2020clearing}
K.-A. Tran, C.~Sakalis, M.~Sj{\"a}lander, A.~Ros, S.~Kaxiras, and A.~Jimborean,
  ``Clearing the shadows: Recovering lost performance for invisible speculative
  execution through hw/sw co-design,'' in \emph{Proceedings of the ACM
  International Conference on Parallel Architectures and Compilation
  Techniques}, 2020, pp. 241--254.

\bibitem{trippel2018meltdownprime}
C.~Trippel, D.~Lustig, and M.~Martonosi, ``Meltdownprime and spectreprime:
  Automatically-synthesized attacks exploiting invalidation-based coherence
  protocols,'' \emph{arXiv preprint arXiv:1802.03802}, 2018.

\bibitem{trippel2017tricheck}
C.~Trippel, Y.~A. Manerkar, D.~Lustig, M.~Pellauer, and M.~Martonosi,
  ``Tricheck: Memory model verification at the trisection of software,
  hardware, and isa,'' \emph{ACM SIGPLAN Notices}, vol.~52, no.~4, pp.
  119--133, 2017.

\bibitem{weisse2019nda}
O.~Weisse, I.~Neal, K.~Loughlin, T.~F. Wenisch, and B.~Kasikci, ``Nda:
  Preventing speculative execution attacks at their source,'' in
  \emph{Proceedings of the 52nd Annual IEEE/ACM International Symposium on
  Microarchitecture}, 2019, pp. 572--586.

\bibitem{yan2018invisispec}
M.~Yan, J.~Choi, D.~Skarlatos, A.~Morrison, C.~Fletcher, and J.~Torrellas,
  ``Invisispec: Making speculative execution invisible in the cache
  hierarchy,'' in \emph{2018 51st Annual IEEE/ACM International Symposium on
  Microarchitecture (MICRO)}.\hskip 1em plus 0.5em minus 0.4em\relax IEEE,
  2018, pp. 428--441.

\bibitem{yu2020speculative}
J.~Yu, N.~Mantri, J.~Torrellas, A.~Morrison, and C.~W. Fletcher, ``Speculative
  data-oblivious execution: Mobilizing safe prediction for safe and efficient
  speculative execution,'' in \emph{2020 ACM/IEEE 47th Annual International
  Symposium on Computer Architecture (ISCA)}.\hskip 1em plus 0.5em minus
  0.4em\relax IEEE, 2020, pp. 707--720.

\bibitem{yu2019speculative}
J.~Yu, M.~Yan, A.~Khyzha, A.~Morrison, J.~Torrellas, and C.~W. Fletcher,
  ``Speculative taint tracking (stt): A comprehensive protection for
  speculatively accessed data,'' in \emph{Proceedings of the 52nd Annual
  IEEE/ACM International Symposium on Microarchitecture}.\hskip 1em plus 0.5em
  minus 0.4em\relax ACM, 2019, pp. 954--968.

\bibitem{zhao2022pinned}
Z.~N. Zhao, H.~Ji, A.~Morrison, D.~Marinov, and J.~Torrellas, ``Pinned loads:
  taming speculative loads in secure processors,'' in \emph{Proceedings of the
  27th ACM International Conference on Architectural Support for Programming
  Languages and Operating Systems}, 2022.

\end{thebibliography}

\end{document}